\documentclass[prl,twocolumn,superscriptaddress,balance]{revtex4-1}
\usepackage{hyperref,color}
\usepackage[svgnames]{xcolor}
\usepackage{amsmath,amssymb,bm}
\usepackage{hyperref}
\usepackage{graphicx}
\usepackage{float}
\usepackage{subfiles}
\usepackage{commath}
\usepackage{cleveref}
\usepackage{balance}
\usepackage[normalem]{ulem}
\usepackage{balance}

\hypersetup{
colorlinks,
citecolor=Salmon,
linkcolor=Teal,
urlcolor=Teal}

\usepackage{pdfpages}
\makeatletter
\AtBeginDocument{\let\LS@rot\@undefined}
\makeatother

\begin{document}

\newcommand\rkk[1]{\textnormal{\color{red!50!black}{[rkk: #1]}}}

\newcommand\AD[1]{\textnormal{\color{blue!50!black}{[AD: #1]}}}

\title{Coexistence of Canted Antiferromagnetism and Bond-order in $\nu=0$ Graphene }
\author{Ankur Das}
\affiliation{Department of Condensed Matter Physics, Weizmann
Institute of Science, Rehovot, 76100 Israel}
\affiliation{Department of Physics \& Astronomy, University of Kentucky,
Lexington, KY 40506, USA}
\author{Ribhu K. Kaul }
\author{Ganpathy Murthy}
\affiliation{Department of Physics \& Astronomy, University of Kentucky,
Lexington, KY 40506, USA}

\begin{abstract}
Motivated by experimental studies of graphene in the quantum Hall regime,
we revisit the phase diagram of a single sheet of graphene at charge
neutrality. Because of spin and valley degeneracies, interactions play a
crucial role in determining the nature of ground state. We show that,
generically, in the regime of interest there is a region of coexistence
between magnetic and bond orders in the phase diagram. We demonstrate this
result both in continuum and lattice models, and argue that the coexistence
phase naturally provides an explanation for unreconciled experimental
observations on the quantum Hall effect in graphene.
\end{abstract}
\maketitle

{\em Introduction:} The quantum Hall effect is a fundamental manifestation
of topology, quantum mechanics and many-particle physics in two dimensions
\cite{PrangeGirvin1990,dassarma1996:qhe}. Discovered originally in
semiconductor heterostructures, it found a new realization in graphene two
decades later \cite{Berger_etal_2004,Novoselov_etal_2004,Zhang_etal_2005, neto2009:rmp}.
Graphene brings several tantalizing twists to the original quantum Hall
problem that arise due to its honeycomb lattice \cite{Zhang_Kim_2005,Geim_Novoselov_2007, neto2009:rmp, DasSarma_etal_2011_RMP}.
At low energies graphene has a
relativistic linear dispersion leading to an effective Dirac equation
near charge neutrality, which leads to a distinct Landau spectrum
\cite{Zheng_Ando_2002,Gusynin_Sharapov_2005}. Additionally, there are two
copies of the Landau levels due to valley degeneracy, causing
electron-electron interactions to play a crucial role in selecting the
ground state  even for integer fillings
\cite{zhang2006:nu0,Checkelsky_Li_Ong_2009,Zhang_etal_2009}.

Here we focus on the ground state at charge neutrality ($\nu=0$), which
corresponds to an  electron count that fills precisely two of the four
(almost) degenerate $n=0$ Landau levels (LLs). We will call this manifold
of states the zero-energy LLs (ZLLs). At the noninteracting level, the
Zeeman energy splits the four degenerate $n=0$ LLs into pairs of two-fold
degenerate ones, picking a fully polarized ground state
\cite{abanin2006:nu0}. Since the ZLLs have equal contributions from
particle-like and hole-like states, at the edge one linear combination
of the valleys has a particle-like dispersion, while the orthogonal linear
combination has a hole-like dispersion. The edge of a fully polarized bulk
state develops a pair of counter-propagating charged chiral modes protected
by spin-rotation symmetry, manifesting the quantum spin Hall effect
\cite{abanin2006:nu0}. The addition of Coulomb interactions gaps the
single-particle electron spectrum everywhere, but preserves the two
gapless counter-propagating charge modes (protected by $S_z$ conservation),
promoting them into a helical Luttinger liquid \cite{brey2006:nu0}.

From pioneering experiments
\cite{zhang2006:nu0,jiang2007:nu0,young2012:nu0,Maher_Kim_etal_2013}, we now
know that the ground state depends on the balance between the orbital
magnetic field, $B_\perp$ (perpendicular to the graphene sheet) and
the total field, $B_{\rm tot}$ (which enters via the Zeeman energy
$E_Z$ and can be tuned by applying an in-plane field). For $E_Z$ less
than a critical value $E_Z^*$ all charge excitations in the bulk and
the edges are completely gapped. However, for $E_Z>E_Z^*$, one obtains
a gapped bulk with a two-terminal edge conductance of (almost) $2e^2/h$
\cite{young2014:nu0}, which is expected of the helical Luttinger liquid.
While the nature of the phase for $E_Z<E_Z^*$ has not been conclusively
identified in experiment, a continuous phase transition to it from the
fully polarized state is observed \cite{young2014:nu0}. Based on a
Hartree-Fock (HF) treatment of a continuum model which keeps only the
ZLLs with ultra-short range interactions, it is believed that the
$E_Z<E_Z^*$ phase is a canted antiferromagnet \cite{kharitonov2012:nu0}.
While this proposal is consistent with recent magnon transmission
experiments \cite{takei2016:nu0,wei2018:science,Stepanov_2018}
that imply that the state is magnetic, it is in tension with STM studies
\cite{li2019:stm} which find evidence for bond order in the $E_Z<E_Z^*$
insulating phase at $\nu=0$.

In this Letter we offer a resolution to this paradox. We propose that
the seemingly contradictory observations arise from the coexistence
of magnetism and bond order at charge neutrality, which was absent in
previous theoretical phase diagrams. We show by HF methods (known to be
reliable in the integer QHE \cite{PrangeGirvin1990}), both in the
continuum and on the lattice, that coexistence is a generic feature in
the regime of interest. In the continuum model, justified at weak
$B_\perp$ relevant to experiment, we first show that  a general HF
analysis in the ZLLs depends only on six couplings constants that
parametrize the electron-electron interactions. We then show that generic
choices of these couplings lead to coexistence. In a complementary, more
microscopic, HF analysis on the lattice in a magnetic field with $1/q$
quanta of flux penetrating each unit cell, we find ubiquitous evidence
for coexistence for small and moderate values of $q$ up to $36$.
Careful extrapolation to large $q$ of our numerical data demonstrates
that the coexistence survives in the $B_\perp$ regime relevant to
experiments (for reference, $B_\perp=10$T gives $q\simeq10000$).
Since coexistence is generically present in both limiting cases,
we argue that it can explain the experimental observations
\cite{young2014:nu0,li2019:stm}, especially since disorder, which
pins the bond-order, will only enhance its presence in the physical
system. 

A microscopic model for graphene in a magnetic field that is expected
to harbor all the phenomena discussed takes the general form,
\begin{equation}
\label{eq:hlatt}
H_{\rm latt} = -\sum_{\langle ij \rangle} t_{ij}c^\dagger_{is} c_{js}-
E_Z\sum_{i s}s c^\dagger_{is} c_{is} +H^{(4)}_{\rm int}
\end{equation}
where $c_{is}$ destroys an electron on the $i^{\rm th}$ site of the
honeycomb lattice with spin $s=\pm 1$. The Zeeman term, $E_Z= g \mu_B
B_{\rm tot}/2$ and the hopping $t_{ij}=t e^{i \int_i^j \vec A .d\vec l
}$ with $\vec A$ chosen so $\nabla \times \vec A = \hat z B_\perp$,
together describe the free part of the Hamiltonian. The magnetic field
introduces the length scale $\ell=\sqrt{\frac{h}{eB}}$, such that an
area of $2\pi\ell^2$ is pierced by one flux quantum. Since for
$B_\perp=1\ T$ $\ell=25\ nm$, it is clear that $\ell\gg a$, where
$a$ is the lattice spacing.  $H^{(4)}_{\rm int}$ is a four-fermi
electron-electron interaction whose precise form is unknown -- we
shall discuss specific forms for it below.

{\em Continuum:} In this limit justified for $\ell\gg a$, one restricts
attention to low-energy states near the $K,K'$ points, linearizing the
band structure to  Dirac equations at each valley. Momentum conservation,
when applied to two-body interactions, forces the conservation of particle
number in the two valleys independently, leading to a $U(1)$ symmetry in the valley
space \cite{alicea2006:gqhe}. An orbital $B$ field is introduced by
minimal coupling into the Dirac equation, leading to four copies (spin
and valley) of a relativistic Landau level spectrum. The interacting
Hamiltonian  projected into the ZLLs is,
\begin{eqnarray}
\label{eq:hcont}
&H_{\rm cont} = -E_Z \sum\limits_{\alpha, k, s} s c^\dagger_{\alpha k s}    c_{\alpha k s} + \sum\limits_{{\bf q}\mu} \frac{v_\mu({\bf q}) :\rho_\mu({\bf q})\rho_\mu(-{\bf q}):}{2L_xL_y}   
\\
&\rho_\mu({\bf q})=\sum\limits_{k,s,\alpha,\beta}e^{-(\frac{q^2}{4}+iq_x(k-\frac{q_y}{2}))\ell^2}c^{\dagger}_{\alpha k-q_y s}\tau_\mu^{\alpha\beta} c_{\beta k s}\nonumber
\end{eqnarray}
where $ c_{\alpha k s} $ destroys an electron with spin $s$ in valley
$\alpha$ and $y$-momentum $k$, and $\tau_\mu$ are Pauli matrices in the
valley space. We work in the Landau gauge $\vec A=(0,B_\perp x)$ on an
$L_x\times L_y$ sample with periodic boundary conditions in $y$.  
Since the valley and sublattice indices are tied in the ZLLs, no sublattice
index appears. The functions $v_\mu({\bf q})$ are the Fourier
transforms of the effective interactions (in the ZLLs) in the $\mu=0,x,y,z$
valley channels ($\tau_0$ is the unit matrix). The $U(1)$ valley symmetry
forces $v_x({\bf q})=v_y({\bf q})$.
The phase diagram of Eq.~(\ref{eq:hcont}) can be calculated in the HF
approximation with the averages $\langle c^\dagger_{\alpha ks
}c_{\alpha^\prime k^\prime s^\prime}\rangle = \delta_{kk^\prime}
\Delta^{ss^\prime}_{\alpha\alpha^\prime}$ preserves translation
invariance up to an inter-valley coherence.  Inter-valley coherence signifies incipient bond-order, though to realize a
bond-ordered state breaking lattice translation symmetries requires physics beyond the continuum model.
Kharitonov \cite{kharitonov2012:nu0} assumed
ultra-short-ranged interactions in real-space ($v_\mu({\bf q})\equiv
v_\mu$ constant), and found a HF phase diagram with four phases: canted
antiferromagnetic (CAF, characterized by the order parameter $Tr(
\tau_z\sigma_x\Delta) \neq0$ and $Tr(\sigma_z\Delta)<2$), fully
polarized (F, characterized by $Tr(\sigma_z\Delta)=2$),
charge-density-wave (CDW, characterized by
$Tr(\tau_z \Delta)\neq0$), and bond-ordered (BO, characterized by
$Tr(\tau_x\Delta)\neq0$). There is no coexistence of order parameters 
in this model, and all transitions except for CAF to  F
are first-order. Experimental graphene samples are believed to be in
the CAF regime for purely perpendicular fields, which needs
$v_x=v_y<0$, and $v_z>|v_x|$. Kharitonov found in his model that 
$E_Z^*=|v_x|/\pi\ell^2$, leading to the conclusion that increasing
$E_Z$ while keeping $B_\perp$ fixed will eventually lead to a fully
polarized bulk state for $E_Z>E_Z^*$ via a second-order phase
transition, consistent with experiment \cite{young2014:nu0}.

We now show that considering a more general form of the interaction by
relaxing the ultra-short-range assumption  leads generically to
coexistence between the canted antiferromagnet and bond-ordered states
near their phase boundary in the ultra-short-range model. While the
functions $v_\mu({\bf q})$ have an infinite number of degrees of freedom,
the translation-invariant HF ground state energy depends only on two
specific numbers for each $v_\mu$: The Hartree coupling $g_{\mu,H} =
\frac{v_\mu({\bf 0})}{2\pi l^2}$ and the Fock coupling $g_{\mu,F} = \int
\frac{d {\bf q}}{(2\pi)^2}v_\mu({\bf q})e^{-q^2l^2/2}$. These six coupling
constants completely characterize the HF energies of all
translation-invariant ground states in the ZLLs. The assumption in previous
work \cite{kharitonov2012:nu0} that the interactions remain short-range
on the lattice scale $a\ll\ell$ even in the effective theory in the ZLLs
forces $g_{\mu,H}=g_{\mu,F}$. As we now show, it is this restrictive
assumption that leads to  the lack of coexistence in the phase
diagram in previous work \cite{kharitonov2012:nu0}. 

In the regime of coupling constants of interest in real graphene samples,
where the ground states are CAF and/or BO, we find that three of the
couplings  $g_{0,H},g_{0,F},g_{z,H}$ play no role in selecting the ground
state. We are left with just three independent couplings $g_{z,F},g_{xy,H},
g_{xy,F}$. We assume an ansatz for the two occupied orbitals that
interpolates between the CAF and the BO states \cite{MSF2014}.
\begin{eqnarray}
\label{two-angle-ansatz}
|a\rangle=&\frac{1}{\sqrt{2}}\big(c_a|{K\uparrow}\rangle-s_a|{K
\downarrow}\rangle+c_a|{K'\uparrow}\rangle+s_a|{K'\downarrow}\rangle\big)\\
|b\rangle=&\frac{1}{\sqrt{2}}\left(-c_b|{K\uparrow}\rangle+s_b|{K
\downarrow}\rangle+c_b|{K'\uparrow}\rangle+s_b|{K'\downarrow}\rangle\right)
\end{eqnarray}
where $c_\alpha=\cos{\frac{\psi_\alpha}{2}}$ and $s_\alpha =
\sin{\frac{\psi_\alpha}{2}}$.  The CAF state corresponds to
$\psi_a=\psi_b=\theta$, the canting angle, and the BO state corresponds
to $\psi_a=0,\ \psi_b=\pi$. In a generic state, these two angles are
independently minimized. We have verified that this ansatz correctly
describes the states of interest by numerically carrying out iterative
HF starting from random ``seed'' $\Delta$-matrices. We find two  necessary
conditions for coexistence: $|g_{xy,F}|>|g_{xy,H}|$ and $E_Z>0$. Fig. 1
shows the order parameters for the BO, CAF and F states as a function
of $E_Z$ for a particular choice of our parameters. With this choice,
the system starts in the BO phase at zero $E_Z$, undergoes a phase
transition to a phase with coexistence between BO and CAF for
intermediate $E_Z$, goes through another transition to a pure CAF phase,
and finally to the F phase. All transitions are second-order. Fig. 2
is a section of the phase diagram at constant $g_{xy,F}=-1, E_Z=1$,
clearly showing that coexistence is absent with the usual ultra-short
range assumption $g_{xy,H}=g_{xy,F}$, but appears when $g_{xy,H}-
g_{xy,F}>0$. Evidently, $g_{xy,H}-g_{xy,F}$ determines the sign of the
energy-energy coupling between the two order parameters
\cite{Bruce_Aharony1975} in a Landau theory of the phase transition.

\begin{figure}[t]
\centering
\includegraphics[width=0.95\columnwidth]{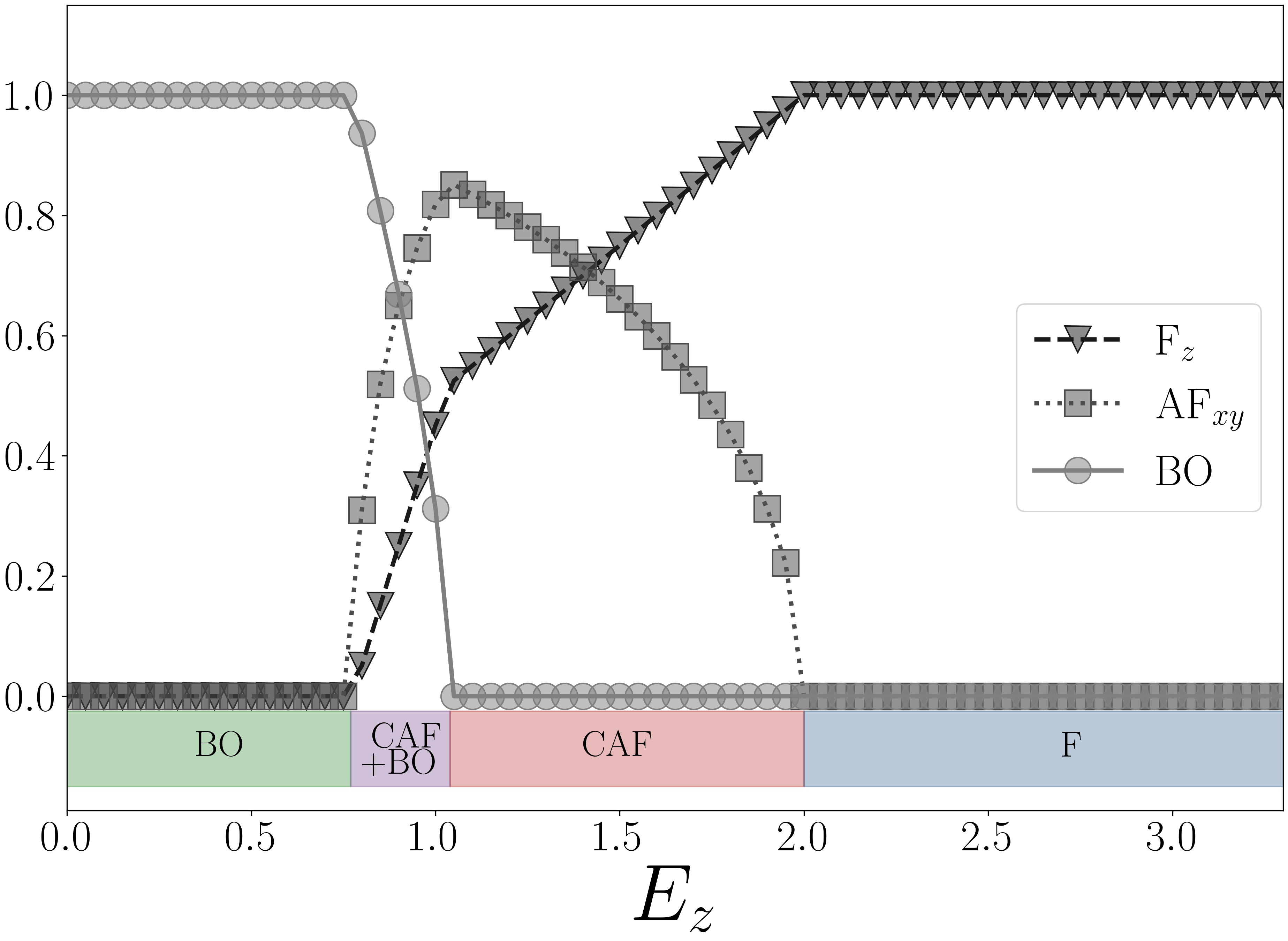}
\caption{
Order parameters obtained from our generalized HF study of the continuum
theory, Eq.~(\ref{eq:hcont}) plotted as a function of the Zeeman energy,
$E_Z$. We have chosen the interaction parameters $g_{z,F}=0.1$,
$g_{xy,H}=-0.75$, $g_{xy,F}=-1$. The bar at the bottom shows the phase
the system is in based on which orders have condensed. For $E_z=0$ the
system is in the BO (bond-ordered) phase. For $E_z$ very large the
system is in (F) ferromagnetic phase. Varying $E_Z$ between these limits,
the system goes through two intermediate phases, a canted anti-ferromagnet
(CAF) without and with bond order coexistent (CAF+BO) (all three order
parameters are non-zero). All the transitions are continuous in our HF
theory.}
\label{fig:zeeman}
\end{figure}

In order for $g_{\mu,H}$ to be significantly different from $g_{\mu,F}$
one needs the relevant function $v_{\mu}({\bf q})$ to vary on the scale
of the magnetic length $\ell$ in real-space and be non-monotonic. While
the renormalization of the effective interactions in the ZLLs from very
high-energy states should be independent of $\ell$, the Dirac-Landau
quantization of energy levels, in combination with LL-mixing induced by
the Coulomb interaction \cite{feshamiFertig2016:nu0}, naturally introduces
this scale into the effective interactions while integrating out lower
energies. We show an explicit model calculation of this effect in the
supplemental material (SM) \cite{SM}.

{\em Lattice:} A canonical way to proceed is to carry out a renormalization
group calculation to determine the effective interactions in the ZLL due to
LL-mixing. Unfortunately, because of the large LL-mixing induced by the
Coulomb interaction \cite{feshamiFertig2016:nu0}, there is no natural small
parameter that would justify such a calculation. We will proceed in another
direction by carrying out a lattice HF calculation in the presence of a
small orbital flux (for the noninteracting limit, see, for example
\cite{Rhim_Park_2012,Das_Kaul_Murthy_2020}) per unit cell
\cite{Jung_Macdonald_2009,Lado_rossier_2014,Lukose_Shankar_2016,Mishra_Hassan_Shankar_2016,Mishra_Hassan_Shankar_2017,Mishra_Lee_2018,Mastropietro_2019,Giuliani_Mastropietro_Porta_2020}.
Since no projection to the low-energy manifold is performed, all LL-mixing
effects are automatically included. Furthermore,  lattice scale physics
(C$_3$ symmetry, reciprocal lattice vectors, etc) that plays an important
role in the bond order is kept fully, while it is absent in the continuum.
This approach also allows us to answer the interesting question of whether
short-range interactions on the lattice can lead to effective interactions in
the ZLL with structure on the scale of $\ell$, at least at the level of HF.  
\begin{figure}[t]
\centering
\includegraphics[width=0.95\columnwidth]{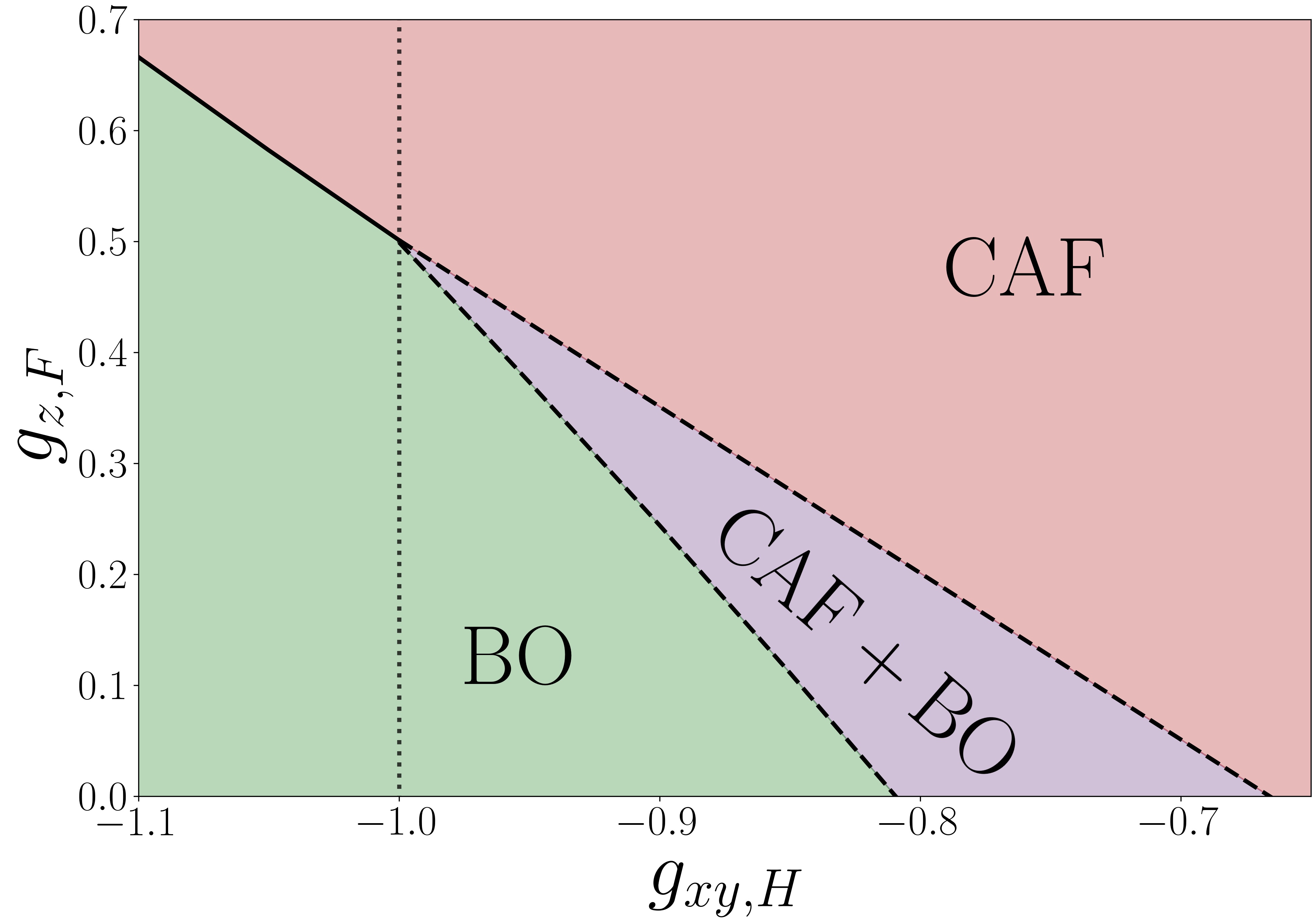}
\caption{A section of the HF phase diagram obtained from our continuum
theory, Eq.~\ref{eq:hcont}. Coexistence between CAF and BO at fixed $E_Z$
can be seen in a robust region. The plots are made for $g_{xy,F}=-1,
E_z=1.0$. Two necessary conditions for coexistence are $0>g_{xy,H}>
g_{xy,H}$ and $E_Z>0$. The ultra-short-range result is the dotted
vertical line at $g_{xy,H}=-1$.}
\label{fig:MLG_phases}
\end{figure}
\begin{figure*}
\centering
\includegraphics[width=0.99\textwidth]{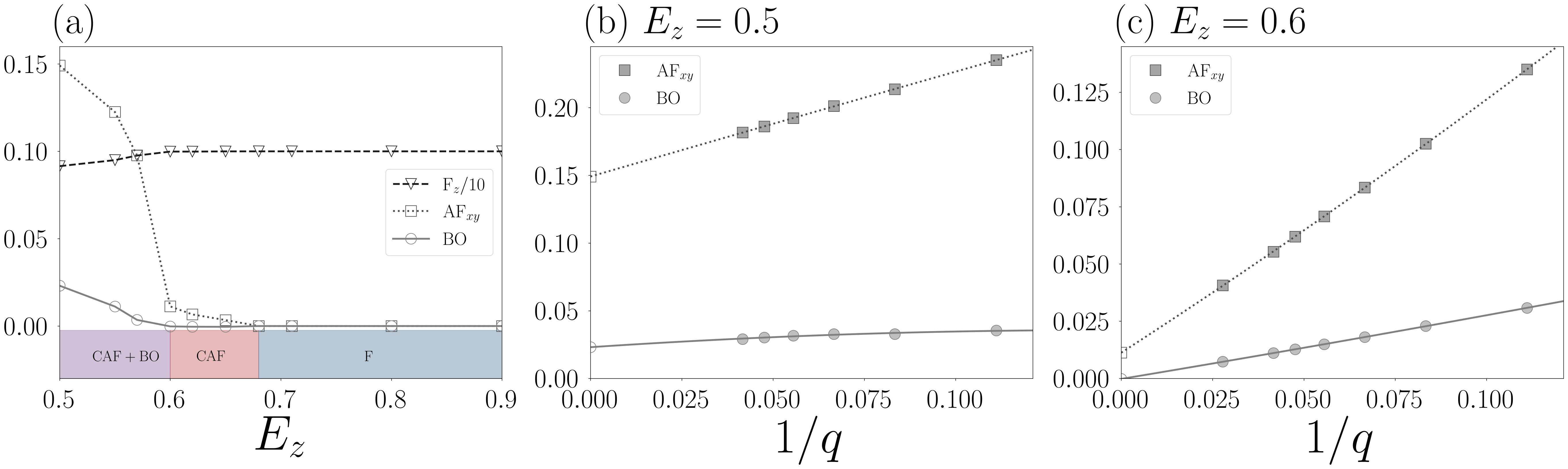}
\caption{Illustrative HF results for the lattice model defined by
Eqs.~(\ref{eq:hlatt}) and~(\ref{eq:lattinteraction}). The numerical
results are obtained on lattices with a flux of $1/q$ quanta per unit
cell and then extrapolated to the weak field  regime ($q\rightarrow
\infty$). (a) The extrapolated order parameters for $g=0.3,~ U=3.5$ as a
function of $E_Z$. Note that there are two phase transitions from zero
to large Zeeman coupling, consistent with our continuum result. The
phases are labeled in the bottom bar. (b,c) Examples of the extrapolations
of the AF$_{xy}$ and BO order parameters to the $q\rightarrow \infty$
limit used to produce (a).  For finite-$q$ coexistence between magnetism
and BO is ubiquitous, but the BO vanishes at intermediate Zeeman as
$q\rightarrow \infty$, resulting in a pure CAF phase for $E_Z>0.6$.}
\label{fig:lattice}
\end{figure*}
We use,
\begin{equation}
H_{int}^{(4)}=\frac{U}{2}\sum\limits_{i} \big({ n}_i
\big)^2-2g\sum\limits_{\langle ij\rangle} {\vec S_i}\cdot {\vec S}_j,
\label{eq:lattinteraction}
\end{equation}
where $n_i=\sum_{s}c^\dagger_{is}c_{is}$ and ${\vec S}_i =\frac{1}{2}
\sum_{s,s^\prime} c^\dagger_{is}{\vec \sigma}_{ss^\prime}c_{is^\prime}$. 
The first term is the  Hubbard interaction, and the
second is a nearest-neighbor Heisenberg spin exchange. We treat this
model in HF approximation allowing for translation symmetry
breaking~\cite{SM}.

As expected, the phase diagram we find is much richer than that found
in the continuum, with several different types of magnetic order and bond
order making their appearance in different ranges of parameters. We focus
on the issue of interest, setting aside the full phase diagram for a
future publication. Computational resources limit us to a maximum $q$
of 36, which corresponds to $B_\perp$ much larger than experimentally
accessible fields. We circumvent this shortcoming by extrapolating our
data to the large-$q$ limit, which corresponds to experimentally
realizable fields. The extrapolated order parameters are shown in Fig.
\ref{fig:lattice}(a) for a particular choice of couplings $U,g$. There
are two distinct phase transitions at $E_{Z1}$ and $E_{Z2}$ as $E_Z$ is
increased. The other two panels show how the extrapolation is done for
representative points in the CAF/BO coexistent phase (b), and for the
pure CAF phase (c). For $E_Z<E_{Z1}$, as shown in Fig.
\ref{fig:lattice}(b), the order parameters of the CAF and BO both
saturate to nonzero values in the limit $q\to\infty$. However, for
$E_{Z1}<E_Z<E_{Z2}$ (Fig. \ref{fig:lattice}(c)) the bond-order vanishes
in the continuum limit, while the CAF saturates to a nonzero value. 

Strikingly, the lattice results, when extrapolated to the  weak-field
limit, show the same sequence of phases with increasing Zeeman coupling
as in the continuum. Even thought the two calculations approach the
problem from opposite limits they converge on the same generic nature
of the coexistence between CAF and BO. As anticipated, even though $U$
and $J$ are ultra-short-range interactions, the LL-mixing inherent in
the full lattice calculation has succeeded in generating structure in
the effective $v_\mu({\bf q})$ on the scale of $\ell$.

{\em Discussion:}  Our theoretical results show that there is no
contradiction between magnon transmission experiments \cite{wei2018:science}
that provide direct evidence of low-energy spin excitations and STM
experiments \cite{li2019:stm} which show evidence for bond-order.
Furthermore they imply that the interactions in charge-neutral graphene
lie in a region of parameter space supporting the coexistence of CAF and
bond-order at low Zeeman coupling, and that the effective interactions
$v_{\mu}({\bf q})$ in the ZLLs necessarily have structure on the scale of the
magnetic length. This raises the possibility that other theoretical
results (for bilayer graphene, for example) based on ultra-short-range
interactions may need to be revisited. For future experiments, we predict
a new phase transition. Indeed, as we have shown bond-order will
decrease as $E_Z$ increases, vanishing continuously at a ``lower''
critical Zeeman coupling $E_{Z1}$. This is distinct from the ``upper''
critical Zeeman field $E_{Z2}$ at which there is a continuous
transition to the fully spin-polarized state. STM experiments carried
out over a range of in-plane fields can  test this
prediction.

An important aspect of the experiment not taken into account in our
studies is the effect of disorder. We generically expect disorder to
enhance bond order, though of course it will have other effects as well \cite{Hong_etal_2021}. While bond-order breaks translational invariance
spontaneously, disorder breaks this symmetry explicitly, favoring the
bond-ordered state over the translation-invariant CAF state. Thus, we
can expect STM experiments to see bond-order over a wider range of $E_Z$
than we found theoretically. While technically, based on the mapping to a
random-field Ising model \cite{Imry_Ma1975,Binder1983,Aizenman_Wehr1989},
one may conclude that long-range bond-order is destroyed by disorder,
this clearly does not have implications for STM experiments, which
measure the local strength of bond-order.

In summary, we have resolved a seeming contradiction in the nature of
the low-Zeeman charge-neutral state of graphene in the quantum Hall
regime. By two complementary methods we find that coexistence between
CAF and BO orders is generic. From the theoretical side, the neighborhood
of the phase transition between the CAF and the BO phases in $\nu=0$
graphene is interesting, because it may host an approximate $SO(5)$
symmetry \cite{Wu_MacDonald_Jolicoeur_2014,Wang_Assaad_etal_2021} and
field theories for this transition contain topological terms
\cite{Sachdev_Lee_2015} which allow certain excitations in either phase
to carry the quantum numbers of the other. These intriguing ideas provide
further  motivation for future experimental and theoretical work on bond
order in $\nu=0$ graphene.

\acknowledgements
The authors acknowledge supported from NSF DMR-1611161, German-Israeli
Foundation (GIF) Grant No. I-1505-303.10/2019, Minerva Foundation, Dean
of Faculty and Israel planning and budgeting committee for financial
support (AD), NSF DMR-2026947 (RKK) and US-Israel Binational Science
Foundation  Grant no. 2016130 (GM). The authors thank Benjamin Sac\'ep\'e,
Alexis Coissard, Adolfo Grushin, and C\'ecile Repellin for stimulating
conversations, and the Aspen Center for Physics (NSF Grant PHY-1607611)
(RKK, GM) where this work was completed. We would also like to thank the
University of Kentucky Center for Computational Sciences and Information
Technology Services Research Computing for their support and use of the
Lipscomb Compute Cluster and associated research computing resources.
\bibliography{hall}

\begin{thebibliography}{47}%
\makeatletter
\providecommand \@ifxundefined [1]{%
 \@ifx{#1\undefined}
}%
\providecommand \@ifnum [1]{%
 \ifnum #1\expandafter \@firstoftwo
 \else \expandafter \@secondoftwo
 \fi
}%
\providecommand \@ifx [1]{%
 \ifx #1\expandafter \@firstoftwo
 \else \expandafter \@secondoftwo
 \fi
}%
\providecommand \natexlab [1]{#1}%
\providecommand \enquote  [1]{``#1''}%
\providecommand \bibnamefont  [1]{#1}%
\providecommand \bibfnamefont [1]{#1}%
\providecommand \citenamefont [1]{#1}%
\providecommand \href@noop [0]{\@secondoftwo}%
\providecommand \href [0]{\begingroup \@sanitize@url \@href}%
\providecommand \@href[1]{\@@startlink{#1}\@@href}%
\providecommand \@@href[1]{\endgroup#1\@@endlink}%
\providecommand \@sanitize@url [0]{\catcode `\\12\catcode `\$12\catcode
  `\&12\catcode `\#12\catcode `\^12\catcode `\_12\catcode `\%12\relax}%
\providecommand \@@startlink[1]{}%
\providecommand \@@endlink[0]{}%
\providecommand \url  [0]{\begingroup\@sanitize@url \@url }%
\providecommand \@url [1]{\endgroup\@href {#1}{\urlprefix }}%
\providecommand \urlprefix  [0]{URL }%
\providecommand \Eprint [0]{\href }%
\providecommand \doibase [0]{http://dx.doi.org/}%
\providecommand \selectlanguage [0]{\@gobble}%
\providecommand \bibinfo  [0]{\@secondoftwo}%
\providecommand \bibfield  [0]{\@secondoftwo}%
\providecommand \translation [1]{[#1]}%
\providecommand \BibitemOpen [0]{}%
\providecommand \bibitemStop [0]{}%
\providecommand \bibitemNoStop [0]{.\EOS\space}%
\providecommand \EOS [0]{\spacefactor3000\relax}%
\providecommand \BibitemShut  [1]{\csname bibitem#1\endcsname}%
\let\auto@bib@innerbib\@empty
\bibitem [{\citenamefont {Prange}\ and\ \citenamefont
  {Girvin}(1990)}]{PrangeGirvin1990}%
  \BibitemOpen
  \bibfield  {author} {\bibinfo {author} {\bibfnamefont {R.}~\bibnamefont
  {Prange}}\ and\ \bibinfo {author} {\bibfnamefont {S.~M.}\ \bibnamefont
  {Girvin}},\ }\href@noop {} {\emph {\bibinfo {title} {The Quantum Hall
  Effect}}}\ (\bibinfo  {publisher} {Springer Verlag},\ \bibinfo {year}
  {1990})\BibitemShut {NoStop}%
\bibitem [{\citenamefont {Das~Sarma}\ and\ \citenamefont
  {Pinczuk}(1996)}]{dassarma1996:qhe}%
  \BibitemOpen
  \bibinfo {editor} {\bibfnamefont {S.}~\bibnamefont {Das~Sarma}}\ and\
  \bibinfo {editor} {\bibfnamefont {A.}~\bibnamefont {Pinczuk}},\ eds.,\
  \href@noop {} {\emph {\bibinfo {title} {Perspectives in Quantum Hall
  Effects}}}\ (\bibinfo  {publisher} {Wiley},\ \bibinfo {year}
  {1996})\BibitemShut {NoStop}%
\bibitem [{\citenamefont {Berger}\ \emph {et~al.}(2004)\citenamefont {Berger},
  \citenamefont {Song}, \citenamefont {Li}, \citenamefont {Li}, \citenamefont
  {Ogbazghi}, \citenamefont {Feng}, \citenamefont {Dai}, \citenamefont
  {Marchenkov}, \citenamefont {Conrad}, \citenamefont {First},\ and\
  \citenamefont {de~Heer}}]{Berger_etal_2004}%
  \BibitemOpen
  \bibfield  {author} {\bibinfo {author} {\bibfnamefont {C.}~\bibnamefont
  {Berger}}, \bibinfo {author} {\bibfnamefont {Z.}~\bibnamefont {Song}},
  \bibinfo {author} {\bibfnamefont {T.}~\bibnamefont {Li}}, \bibinfo {author}
  {\bibfnamefont {X.}~\bibnamefont {Li}}, \bibinfo {author} {\bibfnamefont
  {A.~Y.}\ \bibnamefont {Ogbazghi}}, \bibinfo {author} {\bibfnamefont
  {R.}~\bibnamefont {Feng}}, \bibinfo {author} {\bibfnamefont {Z.}~\bibnamefont
  {Dai}}, \bibinfo {author} {\bibfnamefont {A.~N.}\ \bibnamefont {Marchenkov}},
  \bibinfo {author} {\bibfnamefont {E.~H.}\ \bibnamefont {Conrad}}, \bibinfo
  {author} {\bibfnamefont {P.~N.}\ \bibnamefont {First}}, \ and\ \bibinfo
  {author} {\bibfnamefont {W.~A.}\ \bibnamefont {de~Heer}},\ }\href {\doibase
  10.1021/jp040650f} {\bibfield  {journal} {\bibinfo  {journal} {The Journal of
  Physical Chemistry B}\ }\textbf {\bibinfo {volume} {108}},\ \bibinfo {pages}
  {19912} (\bibinfo {year} {2004})}\BibitemShut {NoStop}%
\bibitem [{\citenamefont {Novoselov}(2004)}]{Novoselov_etal_2004}%
  \BibitemOpen
  \bibfield  {author} {\bibinfo {author} {\bibfnamefont {K.~S.}\ \bibnamefont
  {Novoselov}},\ }\href {\doibase 10.1126/science.1102896} {\bibfield
  {journal} {\bibinfo  {journal} {Science}\ }\textbf {\bibinfo {volume}
  {306}},\ \bibinfo {pages} {666} (\bibinfo {year} {2004})}\BibitemShut
  {NoStop}%
\bibitem [{\citenamefont {Zhang}\ \emph
  {et~al.}(2005{\natexlab{a}})\citenamefont {Zhang}, \citenamefont {Small},
  \citenamefont {Amori},\ and\ \citenamefont {Kim}}]{Zhang_etal_2005}%
  \BibitemOpen
  \bibfield  {author} {\bibinfo {author} {\bibfnamefont {Y.}~\bibnamefont
  {Zhang}}, \bibinfo {author} {\bibfnamefont {J.~P.}\ \bibnamefont {Small}},
  \bibinfo {author} {\bibfnamefont {M.~E.~S.}\ \bibnamefont {Amori}}, \ and\
  \bibinfo {author} {\bibfnamefont {P.}~\bibnamefont {Kim}},\ }\href {\doibase
  10.1103/PhysRevLett.94.176803} {\bibfield  {journal} {\bibinfo  {journal}
  {Phys. Rev. Lett.}\ }\textbf {\bibinfo {volume} {94}},\ \bibinfo {pages}
  {176803} (\bibinfo {year} {2005}{\natexlab{a}})}\BibitemShut {NoStop}%
\bibitem [{\citenamefont {Castro~Neto}\ \emph {et~al.}(2009)\citenamefont
  {Castro~Neto}, \citenamefont {Guinea}, \citenamefont {Peres}, \citenamefont
  {Novoselov},\ and\ \citenamefont {Geim}}]{neto2009:rmp}%
  \BibitemOpen
  \bibfield  {author} {\bibinfo {author} {\bibfnamefont {A.~H.}\ \bibnamefont
  {Castro~Neto}}, \bibinfo {author} {\bibfnamefont {F.}~\bibnamefont {Guinea}},
  \bibinfo {author} {\bibfnamefont {N.~M.~R.}\ \bibnamefont {Peres}}, \bibinfo
  {author} {\bibfnamefont {K.~S.}\ \bibnamefont {Novoselov}}, \ and\ \bibinfo
  {author} {\bibfnamefont {A.~K.}\ \bibnamefont {Geim}},\ }\href {\doibase
  10.1103/RevModPhys.81.109} {\bibfield  {journal} {\bibinfo  {journal} {Rev.
  Mod. Phys.}\ }\textbf {\bibinfo {volume} {81}},\ \bibinfo {pages} {109}
  (\bibinfo {year} {2009})}\BibitemShut {NoStop}%
\bibitem [{\citenamefont {Zhang}\ \emph
  {et~al.}(2005{\natexlab{b}})\citenamefont {Zhang}, \citenamefont {Tan},
  \citenamefont {Stormer},\ and\ \citenamefont {Kim}}]{Zhang_Kim_2005}%
  \BibitemOpen
  \bibfield  {author} {\bibinfo {author} {\bibfnamefont {Y.}~\bibnamefont
  {Zhang}}, \bibinfo {author} {\bibfnamefont {Y.-W.}\ \bibnamefont {Tan}},
  \bibinfo {author} {\bibfnamefont {H.~L.}\ \bibnamefont {Stormer}}, \ and\
  \bibinfo {author} {\bibfnamefont {P.}~\bibnamefont {Kim}},\ }\href {\doibase
  10.1038/nature04235} {\bibfield  {journal} {\bibinfo  {journal} {Nature}\
  }\textbf {\bibinfo {volume} {438}},\ \bibinfo {pages} {201} (\bibinfo {year}
  {2005}{\natexlab{b}})}\BibitemShut {NoStop}%
\bibitem [{\citenamefont {Geim}\ and\ \citenamefont
  {Novoselov}(2007)}]{Geim_Novoselov_2007}%
  \BibitemOpen
  \bibfield  {author} {\bibinfo {author} {\bibfnamefont {A.~K.}\ \bibnamefont
  {Geim}}\ and\ \bibinfo {author} {\bibfnamefont {K.~S.}\ \bibnamefont
  {Novoselov}},\ }\href {\doibase 10.1038/nmat1849} {\bibfield  {journal}
  {\bibinfo  {journal} {Nature Materials}\ }\textbf {\bibinfo {volume} {6}},\
  \bibinfo {pages} {183} (\bibinfo {year} {2007})}\BibitemShut {NoStop}%
\bibitem [{\citenamefont {Das~Sarma}\ \emph {et~al.}(2011)\citenamefont
  {Das~Sarma}, \citenamefont {Adam}, \citenamefont {Hwang},\ and\ \citenamefont
  {Rossi}}]{DasSarma_etal_2011_RMP}%
  \BibitemOpen
  \bibfield  {author} {\bibinfo {author} {\bibfnamefont {S.}~\bibnamefont
  {Das~Sarma}}, \bibinfo {author} {\bibfnamefont {S.}~\bibnamefont {Adam}},
  \bibinfo {author} {\bibfnamefont {E.~H.}\ \bibnamefont {Hwang}}, \ and\
  \bibinfo {author} {\bibfnamefont {E.}~\bibnamefont {Rossi}},\ }\href
  {\doibase 10.1103/RevModPhys.83.407} {\bibfield  {journal} {\bibinfo
  {journal} {Rev. Mod. Phys.}\ }\textbf {\bibinfo {volume} {83}},\ \bibinfo
  {pages} {407} (\bibinfo {year} {2011})}\BibitemShut {NoStop}%
\bibitem [{\citenamefont {Zheng}\ and\ \citenamefont
  {Ando}(2002)}]{Zheng_Ando_2002}%
  \BibitemOpen
  \bibfield  {author} {\bibinfo {author} {\bibfnamefont {Y.}~\bibnamefont
  {Zheng}}\ and\ \bibinfo {author} {\bibfnamefont {T.}~\bibnamefont {Ando}},\
  }\href {\doibase 10.1103/PhysRevB.65.245420} {\bibfield  {journal} {\bibinfo
  {journal} {Phys. Rev. B}\ }\textbf {\bibinfo {volume} {65}},\ \bibinfo
  {pages} {245420} (\bibinfo {year} {2002})}\BibitemShut {NoStop}%
\bibitem [{\citenamefont {Gusynin}\ and\ \citenamefont
  {Sharapov}(2005)}]{Gusynin_Sharapov_2005}%
  \BibitemOpen
  \bibfield  {author} {\bibinfo {author} {\bibfnamefont {V.~P.}\ \bibnamefont
  {Gusynin}}\ and\ \bibinfo {author} {\bibfnamefont {S.~G.}\ \bibnamefont
  {Sharapov}},\ }\href {\doibase 10.1103/PhysRevLett.95.146801} {\bibfield
  {journal} {\bibinfo  {journal} {Phys. Rev. Lett.}\ }\textbf {\bibinfo
  {volume} {95}},\ \bibinfo {pages} {146801} (\bibinfo {year}
  {2005})}\BibitemShut {NoStop}%
\bibitem [{\citenamefont {Zhang}\ \emph {et~al.}(2006)\citenamefont {Zhang},
  \citenamefont {Jiang}, \citenamefont {Small}, \citenamefont {Purewal},
  \citenamefont {Tan}, \citenamefont {Fazlollahi}, \citenamefont {Chudow},
  \citenamefont {Jaszczak}, \citenamefont {Stormer},\ and\ \citenamefont
  {Kim}}]{zhang2006:nu0}%
  \BibitemOpen
  \bibfield  {author} {\bibinfo {author} {\bibfnamefont {Y.}~\bibnamefont
  {Zhang}}, \bibinfo {author} {\bibfnamefont {Z.}~\bibnamefont {Jiang}},
  \bibinfo {author} {\bibfnamefont {J.~P.}\ \bibnamefont {Small}}, \bibinfo
  {author} {\bibfnamefont {M.~S.}\ \bibnamefont {Purewal}}, \bibinfo {author}
  {\bibfnamefont {Y.-W.}\ \bibnamefont {Tan}}, \bibinfo {author} {\bibfnamefont
  {M.}~\bibnamefont {Fazlollahi}}, \bibinfo {author} {\bibfnamefont {J.~D.}\
  \bibnamefont {Chudow}}, \bibinfo {author} {\bibfnamefont {J.~A.}\
  \bibnamefont {Jaszczak}}, \bibinfo {author} {\bibfnamefont {H.~L.}\
  \bibnamefont {Stormer}}, \ and\ \bibinfo {author} {\bibfnamefont
  {P.}~\bibnamefont {Kim}},\ }\href {\doibase 10.1103/PhysRevLett.96.136806}
  {\bibfield  {journal} {\bibinfo  {journal} {Phys. Rev. Lett.}\ }\textbf
  {\bibinfo {volume} {96}},\ \bibinfo {pages} {136806} (\bibinfo {year}
  {2006})}\BibitemShut {NoStop}%
\bibitem [{\citenamefont {Checkelsky}\ \emph {et~al.}(2009)\citenamefont
  {Checkelsky}, \citenamefont {Li},\ and\ \citenamefont
  {Ong}}]{Checkelsky_Li_Ong_2009}%
  \BibitemOpen
  \bibfield  {author} {\bibinfo {author} {\bibfnamefont {J.~G.}\ \bibnamefont
  {Checkelsky}}, \bibinfo {author} {\bibfnamefont {L.}~\bibnamefont {Li}}, \
  and\ \bibinfo {author} {\bibfnamefont {N.~P.}\ \bibnamefont {Ong}},\ }\href
  {\doibase 10.1103/PhysRevB.79.115434} {\bibfield  {journal} {\bibinfo
  {journal} {Phys. Rev. B}\ }\textbf {\bibinfo {volume} {79}},\ \bibinfo
  {pages} {115434} (\bibinfo {year} {2009})}\BibitemShut {NoStop}%
\bibitem [{\citenamefont {Zhang}\ \emph {et~al.}(2009)\citenamefont {Zhang},
  \citenamefont {Camacho}, \citenamefont {Cao}, \citenamefont {Chen},
  \citenamefont {Khodas}, \citenamefont {Kharzeev}, \citenamefont {Tsvelik},
  \citenamefont {Valla},\ and\ \citenamefont {Zaliznyak}}]{Zhang_etal_2009}%
  \BibitemOpen
  \bibfield  {author} {\bibinfo {author} {\bibfnamefont {L.}~\bibnamefont
  {Zhang}}, \bibinfo {author} {\bibfnamefont {J.}~\bibnamefont {Camacho}},
  \bibinfo {author} {\bibfnamefont {H.}~\bibnamefont {Cao}}, \bibinfo {author}
  {\bibfnamefont {Y.~P.}\ \bibnamefont {Chen}}, \bibinfo {author}
  {\bibfnamefont {M.}~\bibnamefont {Khodas}}, \bibinfo {author} {\bibfnamefont
  {D.~E.}\ \bibnamefont {Kharzeev}}, \bibinfo {author} {\bibfnamefont {A.~M.}\
  \bibnamefont {Tsvelik}}, \bibinfo {author} {\bibfnamefont {T.}~\bibnamefont
  {Valla}}, \ and\ \bibinfo {author} {\bibfnamefont {I.~A.}\ \bibnamefont
  {Zaliznyak}},\ }\href {\doibase 10.1103/PhysRevB.80.241412} {\bibfield
  {journal} {\bibinfo  {journal} {Phys. Rev. B}\ }\textbf {\bibinfo {volume}
  {80}},\ \bibinfo {pages} {241412} (\bibinfo {year} {2009})}\BibitemShut
  {NoStop}%
\bibitem [{\citenamefont {Abanin}\ \emph {et~al.}(2006)\citenamefont {Abanin},
  \citenamefont {Lee},\ and\ \citenamefont {Levitov}}]{abanin2006:nu0}%
  \BibitemOpen
  \bibfield  {author} {\bibinfo {author} {\bibfnamefont {D.~A.}\ \bibnamefont
  {Abanin}}, \bibinfo {author} {\bibfnamefont {P.~A.}\ \bibnamefont {Lee}}, \
  and\ \bibinfo {author} {\bibfnamefont {L.~S.}\ \bibnamefont {Levitov}},\
  }\href {\doibase 10.1103/PhysRevLett.96.176803} {\bibfield  {journal}
  {\bibinfo  {journal} {Phys. Rev. Lett.}\ }\textbf {\bibinfo {volume} {96}},\
  \bibinfo {pages} {176803} (\bibinfo {year} {2006})}\BibitemShut {NoStop}%
\bibitem [{\citenamefont {Brey}\ and\ \citenamefont
  {Fertig}(2006)}]{brey2006:nu0}%
  \BibitemOpen
  \bibfield  {author} {\bibinfo {author} {\bibfnamefont {L.}~\bibnamefont
  {Brey}}\ and\ \bibinfo {author} {\bibfnamefont {H.~A.}\ \bibnamefont
  {Fertig}},\ }\href {\doibase 10.1103/physrevb.73.195408} {\bibfield
  {journal} {\bibinfo  {journal} {Physical Review B}\ }\textbf {\bibinfo
  {volume} {73}} (\bibinfo {year} {2006}),\
  10.1103/physrevb.73.195408}\BibitemShut {NoStop}%
\bibitem [{\citenamefont {Jiang}\ \emph {et~al.}(2007)\citenamefont {Jiang},
  \citenamefont {Zhang}, \citenamefont {Stormer},\ and\ \citenamefont
  {Kim}}]{jiang2007:nu0}%
  \BibitemOpen
  \bibfield  {author} {\bibinfo {author} {\bibfnamefont {Z.}~\bibnamefont
  {Jiang}}, \bibinfo {author} {\bibfnamefont {Y.}~\bibnamefont {Zhang}},
  \bibinfo {author} {\bibfnamefont {H.~L.}\ \bibnamefont {Stormer}}, \ and\
  \bibinfo {author} {\bibfnamefont {P.}~\bibnamefont {Kim}},\ }\href {\doibase
  10.1103/PhysRevLett.99.106802} {\bibfield  {journal} {\bibinfo  {journal}
  {Phys. Rev. Lett.}\ }\textbf {\bibinfo {volume} {99}},\ \bibinfo {pages}
  {106802} (\bibinfo {year} {2007})}\BibitemShut {NoStop}%
\bibitem [{\citenamefont {Young}\ \emph {et~al.}(2012)\citenamefont {Young},
  \citenamefont {Dean}, \citenamefont {Wang}, \citenamefont {Ren},
  \citenamefont {Cadden-Zimansky}, \citenamefont {Watanabe}, \citenamefont
  {Taniguchi}, \citenamefont {Hone}, \citenamefont {Shepard},\ and\
  \citenamefont {Kim}}]{young2012:nu0}%
  \BibitemOpen
  \bibfield  {author} {\bibinfo {author} {\bibfnamefont {A.~F.}\ \bibnamefont
  {Young}}, \bibinfo {author} {\bibfnamefont {C.~R.}\ \bibnamefont {Dean}},
  \bibinfo {author} {\bibfnamefont {L.}~\bibnamefont {Wang}}, \bibinfo {author}
  {\bibfnamefont {H.}~\bibnamefont {Ren}}, \bibinfo {author} {\bibfnamefont
  {P.}~\bibnamefont {Cadden-Zimansky}}, \bibinfo {author} {\bibfnamefont
  {K.}~\bibnamefont {Watanabe}}, \bibinfo {author} {\bibfnamefont
  {T.}~\bibnamefont {Taniguchi}}, \bibinfo {author} {\bibfnamefont
  {J.}~\bibnamefont {Hone}}, \bibinfo {author} {\bibfnamefont {K.~L.}\
  \bibnamefont {Shepard}}, \ and\ \bibinfo {author} {\bibfnamefont
  {P.}~\bibnamefont {Kim}},\ }\href {\doibase 10.1038/nphys2307} {\bibfield
  {journal} {\bibinfo  {journal} {Nature Physics}\ }\textbf {\bibinfo {volume}
  {8}},\ \bibinfo {pages} {550–556} (\bibinfo {year} {2012})}\BibitemShut
  {NoStop}%
\bibitem [{\citenamefont {Maher}\ \emph {et~al.}(2013)\citenamefont {Maher},
  \citenamefont {Dean}, \citenamefont {Young}, \citenamefont {Taniguchi},
  \citenamefont {Watanabe}, \citenamefont {Shepard}, \citenamefont {Hone},\
  and\ \citenamefont {Kim}}]{Maher_Kim_etal_2013}%
  \BibitemOpen
  \bibfield  {author} {\bibinfo {author} {\bibfnamefont {P.}~\bibnamefont
  {Maher}}, \bibinfo {author} {\bibfnamefont {C.~R.}\ \bibnamefont {Dean}},
  \bibinfo {author} {\bibfnamefont {A.~F.}\ \bibnamefont {Young}}, \bibinfo
  {author} {\bibfnamefont {T.}~\bibnamefont {Taniguchi}}, \bibinfo {author}
  {\bibfnamefont {K.}~\bibnamefont {Watanabe}}, \bibinfo {author}
  {\bibfnamefont {K.~L.}\ \bibnamefont {Shepard}}, \bibinfo {author}
  {\bibfnamefont {J.}~\bibnamefont {Hone}}, \ and\ \bibinfo {author}
  {\bibfnamefont {P.}~\bibnamefont {Kim}},\ }\href {\doibase 10.1038/nphys2528}
  {\bibfield  {journal} {\bibinfo  {journal} {Nature Physics}\ }\textbf
  {\bibinfo {volume} {9}},\ \bibinfo {pages} {154–158} (\bibinfo {year}
  {2013})}\BibitemShut {NoStop}%
\bibitem [{\citenamefont {Young}\ \emph {et~al.}(2014)\citenamefont {Young},
  \citenamefont {Sanchez-Yamagishi}, \citenamefont {Hunt}, \citenamefont
  {Choi}, \citenamefont {Watanabe}, \citenamefont {Taniguchi}, \citenamefont
  {Ashoori},\ and\ \citenamefont {Jarillo-Herrero}}]{young2014:nu0}%
  \BibitemOpen
  \bibfield  {author} {\bibinfo {author} {\bibfnamefont {A.~F.}\ \bibnamefont
  {Young}}, \bibinfo {author} {\bibfnamefont {J.~D.}\ \bibnamefont
  {Sanchez-Yamagishi}}, \bibinfo {author} {\bibfnamefont {B.}~\bibnamefont
  {Hunt}}, \bibinfo {author} {\bibfnamefont {S.~H.}\ \bibnamefont {Choi}},
  \bibinfo {author} {\bibfnamefont {K.}~\bibnamefont {Watanabe}}, \bibinfo
  {author} {\bibfnamefont {T.}~\bibnamefont {Taniguchi}}, \bibinfo {author}
  {\bibfnamefont {R.~C.}\ \bibnamefont {Ashoori}}, \ and\ \bibinfo {author}
  {\bibfnamefont {P.}~\bibnamefont {Jarillo-Herrero}},\ }\href {\doibase
  10.1038/nature12800} {\bibfield  {journal} {\bibinfo  {journal} {Nature}\
  }\textbf {\bibinfo {volume} {505}},\ \bibinfo {pages} {528} (\bibinfo {year}
  {2014})}\BibitemShut {NoStop}%
\bibitem [{\citenamefont {Kharitonov}(2012)}]{kharitonov2012:nu0}%
  \BibitemOpen
  \bibfield  {author} {\bibinfo {author} {\bibfnamefont {M.}~\bibnamefont
  {Kharitonov}},\ }\href {\doibase 10.1103/PhysRevB.85.155439} {\bibfield
  {journal} {\bibinfo  {journal} {Phys. Rev. B}\ }\textbf {\bibinfo {volume}
  {85}},\ \bibinfo {pages} {155439} (\bibinfo {year} {2012})}\BibitemShut
  {NoStop}%
\bibitem [{\citenamefont {Takei}\ \emph {et~al.}(2016)\citenamefont {Takei},
  \citenamefont {Yacoby}, \citenamefont {Halperin},\ and\ \citenamefont
  {Tserkovnyak}}]{takei2016:nu0}%
  \BibitemOpen
  \bibfield  {author} {\bibinfo {author} {\bibfnamefont {S.}~\bibnamefont
  {Takei}}, \bibinfo {author} {\bibfnamefont {A.}~\bibnamefont {Yacoby}},
  \bibinfo {author} {\bibfnamefont {B.~I.}\ \bibnamefont {Halperin}}, \ and\
  \bibinfo {author} {\bibfnamefont {Y.}~\bibnamefont {Tserkovnyak}},\ }\href
  {\doibase 10.1103/PhysRevLett.116.216801} {\bibfield  {journal} {\bibinfo
  {journal} {Phys. Rev. Lett.}\ }\textbf {\bibinfo {volume} {116}},\ \bibinfo
  {pages} {216801} (\bibinfo {year} {2016})}\BibitemShut {NoStop}%
\bibitem [{\citenamefont {Wei}\ \emph {et~al.}(2018)\citenamefont {Wei},
  \citenamefont {van~der Sar}, \citenamefont {Lee}, \citenamefont {Watanabe},
  \citenamefont {Taniguchi}, \citenamefont {Halperin},\ and\ \citenamefont
  {Yacoby}}]{wei2018:science}%
  \BibitemOpen
  \bibfield  {author} {\bibinfo {author} {\bibfnamefont {D.~S.}\ \bibnamefont
  {Wei}}, \bibinfo {author} {\bibfnamefont {T.}~\bibnamefont {van~der Sar}},
  \bibinfo {author} {\bibfnamefont {S.~H.}\ \bibnamefont {Lee}}, \bibinfo
  {author} {\bibfnamefont {K.}~\bibnamefont {Watanabe}}, \bibinfo {author}
  {\bibfnamefont {T.}~\bibnamefont {Taniguchi}}, \bibinfo {author}
  {\bibfnamefont {B.~I.}\ \bibnamefont {Halperin}}, \ and\ \bibinfo {author}
  {\bibfnamefont {A.}~\bibnamefont {Yacoby}},\ }\href {\doibase
  10.1126/science.aar4061} {\bibfield  {journal} {\bibinfo  {journal}
  {Science}\ }\textbf {\bibinfo {volume} {362}},\ \bibinfo {pages} {229–233}
  (\bibinfo {year} {2018})}\BibitemShut {NoStop}%
\bibitem [{\citenamefont {Stepanov}\ \emph {et~al.}(2018)\citenamefont
  {Stepanov}, \citenamefont {Che}, \citenamefont {Shcherbakov}, \citenamefont
  {Yang}, \citenamefont {Chen}, \citenamefont {Thilahar}, \citenamefont
  {Voigt}, \citenamefont {Bockrath}, \citenamefont {Smirnov}, \citenamefont
  {Watanabe},\ and\ \citenamefont {et~al.}}]{Stepanov_2018}%
  \BibitemOpen
  \bibfield  {author} {\bibinfo {author} {\bibfnamefont {P.}~\bibnamefont
  {Stepanov}}, \bibinfo {author} {\bibfnamefont {S.}~\bibnamefont {Che}},
  \bibinfo {author} {\bibfnamefont {D.}~\bibnamefont {Shcherbakov}}, \bibinfo
  {author} {\bibfnamefont {J.}~\bibnamefont {Yang}}, \bibinfo {author}
  {\bibfnamefont {R.}~\bibnamefont {Chen}}, \bibinfo {author} {\bibfnamefont
  {K.}~\bibnamefont {Thilahar}}, \bibinfo {author} {\bibfnamefont
  {G.}~\bibnamefont {Voigt}}, \bibinfo {author} {\bibfnamefont {M.~W.}\
  \bibnamefont {Bockrath}}, \bibinfo {author} {\bibfnamefont {D.}~\bibnamefont
  {Smirnov}}, \bibinfo {author} {\bibfnamefont {K.}~\bibnamefont {Watanabe}}, \
  and\ \bibinfo {author} {\bibnamefont {et~al.}},\ }\href {\doibase
  10.1038/s41567-018-0161-5} {\bibfield  {journal} {\bibinfo  {journal} {Nature
  Physics}\ }\textbf {\bibinfo {volume} {14}},\ \bibinfo {pages} {907–911}
  (\bibinfo {year} {2018})}\BibitemShut {NoStop}%
\bibitem [{\citenamefont {Li}\ \emph {et~al.}(2019)\citenamefont {Li},
  \citenamefont {Zhang}, \citenamefont {Yin},\ and\ \citenamefont
  {He}}]{li2019:stm}%
  \BibitemOpen
  \bibfield  {author} {\bibinfo {author} {\bibfnamefont {S.-Y.}\ \bibnamefont
  {Li}}, \bibinfo {author} {\bibfnamefont {Y.}~\bibnamefont {Zhang}}, \bibinfo
  {author} {\bibfnamefont {L.-J.}\ \bibnamefont {Yin}}, \ and\ \bibinfo
  {author} {\bibfnamefont {L.}~\bibnamefont {He}},\ }\href {\doibase
  10.1103/PhysRevB.100.085437} {\bibfield  {journal} {\bibinfo  {journal}
  {Phys. Rev. B}\ }\textbf {\bibinfo {volume} {100}},\ \bibinfo {pages}
  {085437} (\bibinfo {year} {2019})}\BibitemShut {NoStop}%
\bibitem [{\citenamefont {Alicea}\ and\ \citenamefont
  {Fisher}(2006)}]{alicea2006:gqhe}%
  \BibitemOpen
  \bibfield  {author} {\bibinfo {author} {\bibfnamefont {J.}~\bibnamefont
  {Alicea}}\ and\ \bibinfo {author} {\bibfnamefont {M.~P.~A.}\ \bibnamefont
  {Fisher}},\ }\href {\doibase 10.1103/PhysRevB.74.075422} {\bibfield
  {journal} {\bibinfo  {journal} {Phys. Rev. B}\ }\textbf {\bibinfo {volume}
  {74}},\ \bibinfo {pages} {075422} (\bibinfo {year} {2006})}\BibitemShut
  {NoStop}%
\bibitem [{\citenamefont {Murthy}\ \emph {et~al.}(2014)\citenamefont {Murthy},
  \citenamefont {Shimshoni},\ and\ \citenamefont {Fertig}}]{MSF2014}%
  \BibitemOpen
  \bibfield  {author} {\bibinfo {author} {\bibfnamefont {G.}~\bibnamefont
  {Murthy}}, \bibinfo {author} {\bibfnamefont {E.}~\bibnamefont {Shimshoni}}, \
  and\ \bibinfo {author} {\bibfnamefont {H.~A.}\ \bibnamefont {Fertig}},\
  }\href {\doibase 10.1103/PhysRevB.90.241410} {\bibfield  {journal} {\bibinfo
  {journal} {Phys. Rev. B}\ }\textbf {\bibinfo {volume} {90}},\ \bibinfo
  {pages} {241410} (\bibinfo {year} {2014})}\BibitemShut {NoStop}%
\bibitem [{\citenamefont {Bruce}\ and\ \citenamefont
  {Aharony}(1975)}]{Bruce_Aharony1975}%
  \BibitemOpen
  \bibfield  {author} {\bibinfo {author} {\bibfnamefont {A.~D.}\ \bibnamefont
  {Bruce}}\ and\ \bibinfo {author} {\bibfnamefont {A.}~\bibnamefont
  {Aharony}},\ }\href {\doibase 10.1103/PhysRevB.11.478} {\bibfield  {journal}
  {\bibinfo  {journal} {Phys. Rev. B}\ }\textbf {\bibinfo {volume} {11}},\
  \bibinfo {pages} {478} (\bibinfo {year} {1975})}\BibitemShut {NoStop}%
\bibitem [{\citenamefont {Feshami}\ and\ \citenamefont
  {Fertig}(2016)}]{feshamiFertig2016:nu0}%
  \BibitemOpen
  \bibfield  {author} {\bibinfo {author} {\bibfnamefont {B.}~\bibnamefont
  {Feshami}}\ and\ \bibinfo {author} {\bibfnamefont {H.~A.}\ \bibnamefont
  {Fertig}},\ }\href {\doibase 10.1103/PhysRevB.94.245435} {\bibfield
  {journal} {\bibinfo  {journal} {Phys. Rev. B}\ }\textbf {\bibinfo {volume}
  {94}},\ \bibinfo {pages} {245435} (\bibinfo {year} {2016})}\BibitemShut
  {NoStop}%
\bibitem [{SM()}]{SM}%
  \BibitemOpen
  \href@noop {} {\enquote {\bibinfo {title} {Supplemental material},}\
  }\BibitemShut {NoStop}%
\bibitem [{\citenamefont {Rhim}\ and\ \citenamefont
  {Park}(2012)}]{Rhim_Park_2012}%
  \BibitemOpen
  \bibfield  {author} {\bibinfo {author} {\bibfnamefont {J.-W.}\ \bibnamefont
  {Rhim}}\ and\ \bibinfo {author} {\bibfnamefont {K.}~\bibnamefont {Park}},\
  }\href {\doibase 10.1103/PhysRevB.86.235411} {\bibfield  {journal} {\bibinfo
  {journal} {Phys. Rev. B}\ }\textbf {\bibinfo {volume} {86}},\ \bibinfo
  {pages} {235411} (\bibinfo {year} {2012})}\BibitemShut {NoStop}%
\bibitem [{\citenamefont {Das}\ \emph {et~al.}(2020)\citenamefont {Das},
  \citenamefont {Kaul},\ and\ \citenamefont {Murthy}}]{Das_Kaul_Murthy_2020}%
  \BibitemOpen
  \bibfield  {author} {\bibinfo {author} {\bibfnamefont {A.}~\bibnamefont
  {Das}}, \bibinfo {author} {\bibfnamefont {R.~K.}\ \bibnamefont {Kaul}}, \
  and\ \bibinfo {author} {\bibfnamefont {G.}~\bibnamefont {Murthy}},\ }\href
  {\doibase 10.1103/PhysRevB.101.165416} {\bibfield  {journal} {\bibinfo
  {journal} {Phys. Rev. B}\ }\textbf {\bibinfo {volume} {101}},\ \bibinfo
  {pages} {165416} (\bibinfo {year} {2020})}\BibitemShut {NoStop}%
\bibitem [{\citenamefont {Jung}\ and\ \citenamefont
  {MacDonald}(2009)}]{Jung_Macdonald_2009}%
  \BibitemOpen
  \bibfield  {author} {\bibinfo {author} {\bibfnamefont {J.}~\bibnamefont
  {Jung}}\ and\ \bibinfo {author} {\bibfnamefont {A.~H.}\ \bibnamefont
  {MacDonald}},\ }\href {\doibase 10.1103/PhysRevB.80.235417} {\bibfield
  {journal} {\bibinfo  {journal} {Phys. Rev. B}\ }\textbf {\bibinfo {volume}
  {80}},\ \bibinfo {pages} {235417} (\bibinfo {year} {2009})}\BibitemShut
  {NoStop}%
\bibitem [{\citenamefont {Lado}\ and\ \citenamefont
  {Fern\'andez-Rossier}(2014)}]{Lado_rossier_2014}%
  \BibitemOpen
  \bibfield  {author} {\bibinfo {author} {\bibfnamefont {J.~L.}\ \bibnamefont
  {Lado}}\ and\ \bibinfo {author} {\bibfnamefont {J.}~\bibnamefont
  {Fern\'andez-Rossier}},\ }\href {\doibase 10.1103/PhysRevB.90.165429}
  {\bibfield  {journal} {\bibinfo  {journal} {Phys. Rev. B}\ }\textbf {\bibinfo
  {volume} {90}},\ \bibinfo {pages} {165429} (\bibinfo {year}
  {2014})}\BibitemShut {NoStop}%
\bibitem [{\citenamefont {Lukose}\ and\ \citenamefont
  {Shankar}(2016)}]{Lukose_Shankar_2016}%
  \BibitemOpen
  \bibfield  {author} {\bibinfo {author} {\bibfnamefont {V.}~\bibnamefont
  {Lukose}}\ and\ \bibinfo {author} {\bibfnamefont {R.}~\bibnamefont
  {Shankar}},\ }\href {\doibase 10.1103/PhysRevB.94.085135} {\bibfield
  {journal} {\bibinfo  {journal} {Phys. Rev. B}\ }\textbf {\bibinfo {volume}
  {94}},\ \bibinfo {pages} {085135} (\bibinfo {year} {2016})}\BibitemShut
  {NoStop}%
\bibitem [{\citenamefont {Mishra}\ \emph {et~al.}(2016)\citenamefont {Mishra},
  \citenamefont {Hassan},\ and\ \citenamefont
  {Shankar}}]{Mishra_Hassan_Shankar_2016}%
  \BibitemOpen
  \bibfield  {author} {\bibinfo {author} {\bibfnamefont {A.}~\bibnamefont
  {Mishra}}, \bibinfo {author} {\bibfnamefont {S.~R.}\ \bibnamefont {Hassan}},
  \ and\ \bibinfo {author} {\bibfnamefont {R.}~\bibnamefont {Shankar}},\ }\href
  {\doibase 10.1103/PhysRevB.93.125134} {\bibfield  {journal} {\bibinfo
  {journal} {Phys. Rev. B}\ }\textbf {\bibinfo {volume} {93}},\ \bibinfo
  {pages} {125134} (\bibinfo {year} {2016})}\BibitemShut {NoStop}%
\bibitem [{\citenamefont {Mishra}\ \emph {et~al.}(2017)\citenamefont {Mishra},
  \citenamefont {Hassan},\ and\ \citenamefont
  {Shankar}}]{Mishra_Hassan_Shankar_2017}%
  \BibitemOpen
  \bibfield  {author} {\bibinfo {author} {\bibfnamefont {A.}~\bibnamefont
  {Mishra}}, \bibinfo {author} {\bibfnamefont {S.~R.}\ \bibnamefont {Hassan}},
  \ and\ \bibinfo {author} {\bibfnamefont {R.}~\bibnamefont {Shankar}},\ }\href
  {\doibase 10.1103/PhysRevB.95.035140} {\bibfield  {journal} {\bibinfo
  {journal} {Phys. Rev. B}\ }\textbf {\bibinfo {volume} {95}},\ \bibinfo
  {pages} {035140} (\bibinfo {year} {2017})}\BibitemShut {NoStop}%
\bibitem [{\citenamefont {Mishra}\ and\ \citenamefont
  {Lee}(2018)}]{Mishra_Lee_2018}%
  \BibitemOpen
  \bibfield  {author} {\bibinfo {author} {\bibfnamefont {A.}~\bibnamefont
  {Mishra}}\ and\ \bibinfo {author} {\bibfnamefont {S.}~\bibnamefont {Lee}},\
  }\href {\doibase 10.1103/PhysRevB.98.235124} {\bibfield  {journal} {\bibinfo
  {journal} {Phys. Rev. B}\ }\textbf {\bibinfo {volume} {98}},\ \bibinfo
  {pages} {235124} (\bibinfo {year} {2018})}\BibitemShut {NoStop}%
\bibitem [{\citenamefont {Mastropietro}(2019)}]{Mastropietro_2019}%
  \BibitemOpen
  \bibfield  {author} {\bibinfo {author} {\bibfnamefont {V.}~\bibnamefont
  {Mastropietro}},\ }\href {\doibase 10.1103/PhysRevB.99.155154} {\bibfield
  {journal} {\bibinfo  {journal} {Phys. Rev. B}\ }\textbf {\bibinfo {volume}
  {99}},\ \bibinfo {pages} {155154} (\bibinfo {year} {2019})}\BibitemShut
  {NoStop}%
\bibitem [{\citenamefont {Giuliani}\ \emph {et~al.}(2020)\citenamefont
  {Giuliani}, \citenamefont {Mastropietro},\ and\ \citenamefont
  {Porta}}]{Giuliani_Mastropietro_Porta_2020}%
  \BibitemOpen
  \bibfield  {author} {\bibinfo {author} {\bibfnamefont {A.}~\bibnamefont
  {Giuliani}}, \bibinfo {author} {\bibfnamefont {V.}~\bibnamefont
  {Mastropietro}}, \ and\ \bibinfo {author} {\bibfnamefont {M.}~\bibnamefont
  {Porta}},\ }\href {\doibase 10.1007/s10955-019-02405-1} {\bibfield  {journal}
  {\bibinfo  {journal} {Journal of Statistical Physics}\ }\textbf {\bibinfo
  {volume} {180}},\ \bibinfo {pages} {332} (\bibinfo {year}
  {2020})}\BibitemShut {NoStop}%
\bibitem [{\citenamefont {Hong}\ \emph {et~al.}(2021)\citenamefont {Hong},
  \citenamefont {Belke}, \citenamefont {Rode}, \citenamefont {Brechtken},\ and\
  \citenamefont {Haug}}]{Hong_etal_2021}%
  \BibitemOpen
  \bibfield  {author} {\bibinfo {author} {\bibfnamefont {S.~J.}\ \bibnamefont
  {Hong}}, \bibinfo {author} {\bibfnamefont {C.}~\bibnamefont {Belke}},
  \bibinfo {author} {\bibfnamefont {J.~C.}\ \bibnamefont {Rode}}, \bibinfo
  {author} {\bibfnamefont {B.}~\bibnamefont {Brechtken}}, \ and\ \bibinfo
  {author} {\bibfnamefont {R.~J.}\ \bibnamefont {Haug}},\ }\href {\doibase
  https://doi.org/10.1016/j.cap.2021.04.001} {\bibfield  {journal} {\bibinfo
  {journal} {Current Applied Physics}\ }\textbf {\bibinfo {volume} {27}},\
  \bibinfo {pages} {25} (\bibinfo {year} {2021})}\BibitemShut {NoStop}%
\bibitem [{\citenamefont {Imry}\ and\ \citenamefont {Ma}(1975)}]{Imry_Ma1975}%
  \BibitemOpen
  \bibfield  {author} {\bibinfo {author} {\bibfnamefont {Y.}~\bibnamefont
  {Imry}}\ and\ \bibinfo {author} {\bibfnamefont {S.-k.}\ \bibnamefont {Ma}},\
  }\href {\doibase 10.1103/PhysRevLett.35.1399} {\bibfield  {journal} {\bibinfo
   {journal} {Phys. Rev. Lett.}\ }\textbf {\bibinfo {volume} {35}},\ \bibinfo
  {pages} {1399} (\bibinfo {year} {1975})}\BibitemShut {NoStop}%
\bibitem [{\citenamefont {Binder}(1983)}]{Binder1983}%
  \BibitemOpen
  \bibfield  {author} {\bibinfo {author} {\bibfnamefont {K.}~\bibnamefont
  {Binder}},\ }\href@noop {} {\bibfield  {journal} {\bibinfo  {journal}
  {Zeitschrift fur Physik B}\ }\textbf {\bibinfo {volume} {50}},\ \bibinfo
  {pages} {343} (\bibinfo {year} {1983})}\BibitemShut {NoStop}%
\bibitem [{\citenamefont {Aizenman}\ and\ \citenamefont
  {Wehr}(1989)}]{Aizenman_Wehr1989}%
  \BibitemOpen
  \bibfield  {author} {\bibinfo {author} {\bibfnamefont {M.}~\bibnamefont
  {Aizenman}}\ and\ \bibinfo {author} {\bibfnamefont {J.}~\bibnamefont
  {Wehr}},\ }\href {\doibase 10.1103/PhysRevLett.62.2503} {\bibfield  {journal}
  {\bibinfo  {journal} {Phys. Rev. Lett.}\ }\textbf {\bibinfo {volume} {62}},\
  \bibinfo {pages} {2503} (\bibinfo {year} {1989})}\BibitemShut {NoStop}%
\bibitem [{\citenamefont {Wu}\ \emph {et~al.}(2014)\citenamefont {Wu},
  \citenamefont {Sodemann}, \citenamefont {Araki}, \citenamefont {MacDonald},\
  and\ \citenamefont {Jolicoeur}}]{Wu_MacDonald_Jolicoeur_2014}%
  \BibitemOpen
  \bibfield  {author} {\bibinfo {author} {\bibfnamefont {F.}~\bibnamefont
  {Wu}}, \bibinfo {author} {\bibfnamefont {I.}~\bibnamefont {Sodemann}},
  \bibinfo {author} {\bibfnamefont {Y.}~\bibnamefont {Araki}}, \bibinfo
  {author} {\bibfnamefont {A.~H.}\ \bibnamefont {MacDonald}}, \ and\ \bibinfo
  {author} {\bibfnamefont {T.}~\bibnamefont {Jolicoeur}},\ }\href {\doibase
  10.1103/PhysRevB.90.235432} {\bibfield  {journal} {\bibinfo  {journal} {Phys.
  Rev. B}\ }\textbf {\bibinfo {volume} {90}},\ \bibinfo {pages} {235432}
  (\bibinfo {year} {2014})}\BibitemShut {NoStop}%
\bibitem [{\citenamefont {Wang}\ \emph {et~al.}(2021)\citenamefont {Wang},
  \citenamefont {Zaletel}, \citenamefont {Mong},\ and\ \citenamefont
  {Assaad}}]{Wang_Assaad_etal_2021}%
  \BibitemOpen
  \bibfield  {author} {\bibinfo {author} {\bibfnamefont {Z.}~\bibnamefont
  {Wang}}, \bibinfo {author} {\bibfnamefont {M.~P.}\ \bibnamefont {Zaletel}},
  \bibinfo {author} {\bibfnamefont {R.~S.~K.}\ \bibnamefont {Mong}}, \ and\
  \bibinfo {author} {\bibfnamefont {F.~F.}\ \bibnamefont {Assaad}},\ }\href
  {\doibase 10.1103/PhysRevLett.126.045701} {\bibfield  {journal} {\bibinfo
  {journal} {Phys. Rev. Lett.}\ }\textbf {\bibinfo {volume} {126}},\ \bibinfo
  {pages} {045701} (\bibinfo {year} {2021})}\BibitemShut {NoStop}%
\bibitem [{\citenamefont {Lee}\ and\ \citenamefont
  {Sachdev}(2015)}]{Sachdev_Lee_2015}%
  \BibitemOpen
  \bibfield  {author} {\bibinfo {author} {\bibfnamefont {J.}~\bibnamefont
  {Lee}}\ and\ \bibinfo {author} {\bibfnamefont {S.}~\bibnamefont {Sachdev}},\
  }\href {\doibase 10.1103/PhysRevLett.114.226801} {\bibfield  {journal}
  {\bibinfo  {journal} {Phys. Rev. Lett.}\ }\textbf {\bibinfo {volume} {114}},\
  \bibinfo {pages} {226801} (\bibinfo {year} {2015})}\BibitemShut {NoStop}%
\end{thebibliography}%

\newpage\newpage

\clearpage

\newpage\newpage



\section*{Supplementary material}

# Supplemental Materials on the coexistense of canted anti-ferromagnetism and bond order at charge neutrality in the $n = 0$ Landau level of graphene

Ankur Das,[1,2] Ribhu K. Kaul,[2] and Ganpathy Murthy[2]

[1]*Department of Condensed Matter Physics, Weizmann Institute of Science, Rehovot, 76100 Israel*
[2]*Department of Physics & Astronomy, University of Kentucky, Lexington, KY 40506, USA*

Here we present a set of calculations and in supplement to the main text.

## SI. INTRODUCTION

In this set of supplemental materials we present the details of the calculations that substantiate the statements made in the main text. In the first section we show how relaxing the ultra-short-range condition on the interaction functions $v_\mu(\mathbf{q})$ leads, in the Hartree-Fock (HF) approximation to two independent constants $g_{\mu,H}, g_{\mu,F}$ for each type of interaction. We also find the HF ground state energy of the two-angle ansatz mentioned in the main text, and the regions of parameter space where canted antiferromagnetic (CAF) order and/or bond order (BO) exists. We also analyze the stability of these regions within our ansatz, and show two necessary conditions for the coexistence of the CAF and BO: $0 > g_{xy,H} > g_{xy,F}$ and $E_Z > 0$.

As mentioned briefly in the main text, this implies that the function $v_{xy}(\mathbf{q})$ must have nontrivial structure on the scale of the magnetic length $\ell$ in real space. Other than the long-range part of the Coulomb interaction (which does not discriminate between different forms of symmetry-breaking in the ZLLs) all other bare interactions are short-range at the scale of the lattice spacing $a$. Since the lattice scale $a$ is much smaller than the magnetic length $\ell = \sqrt{\frac{\hbar}{eB}}$, any real-space structure in $v_{xy}$ at the length scale $\ell$ must arise from Landau-level (LL) mixing which leads to the renormalization of the bare interaction. In the second section SIII we show an illustrative calculation that uses LL-mixing to renormalize the effective interactions in the $n = 0$ manifold of LLs, which constitutes the Hilbert space for our continuum model. The result is that, generically, such renormalizations do bring in the scale $\ell$ into the effective interactions. Of course, this does not prove that the conditions for coexistence are met in real samples, but it demonstrates a natural mechanism for $v_\mu(\mathbf{q})$ to develop structure on the length scale $\ell$.

In the final section SIV we switch gears to the lattice model and provide details of the gauge chosen for $1/q$ flux quanta going through each primordial graphene unit cell, the types of order allowed in our lattice HF calculation, and show a sample phase diagram for $q = 6$. The ubiquity of bond order and it coexistence with other types of order is evident in this phase diagram.

## SII. HAMILTONIAN WITH NONZERO-RANGE INTERACTIONS

In the monolayer Graphene (MLG) at $\nu = 0$ we have,

$$H_{\text{cont}} = -E_Z \sum_{\alpha ks} s c^\dagger_{\alpha ks} c_{\alpha ks} + \sum_{\substack{k_1 k_2 \\ \mathbf{q} \mu}} \frac{e^{-\ell^2\left(\frac{q^2}{2} + i q_x(k_1 - k_2 - q_y)\right)}}{2 L_x L_y}$$
$$\times v_\mu(\mathbf{q}) c^\dagger_{\alpha s_1 k_1 - q_y} \tau^\mu_{\alpha\alpha'} c_{\alpha' s_1 k_1} c^\dagger_{\beta s_2 k_2 + q_y} \tau^\mu_{\beta\beta'} c_{\beta' s_2 k_2} \quad (S1)$$

The indices $\alpha, \beta$ on the electron creation and destruction operators denote the valley ($\alpha = 0 \Rightarrow K$, $\alpha = 1 \Rightarrow K'$), and the indices $s_i$ denote spin. Since our Hilbert space consists only of $n = 0$ LL states, no LL index appears on the operators. As mentioned in the main text, sublattice-valley locking in the $n = 0$ LL means that the valley index is sufficient to fully describe the states in the ZLLs. Lattice translation symmetry leads to an emergent low-energy $U(1)$ symmetry of separate conservation of the number of particles in each valley, leading to $v_x(\mathbf{q}) = v_y(\mathbf{q})$.

Imposing periodic boundary conditions in the $y$-direction with length $L_y$ we find the guiding centers are

$$k = \frac{2\pi j}{L_y} \ , \ j \in \mathbb{Z} \quad (S2)$$

In the HF approximation, we try to minimize the ground state energy in the class of single Slater determinantal (SSD) states. For an $N$-particle system any such state can be written as

$$|SSD\rangle = \prod_{i=1}^{N} c_i^\dagger |0\rangle \quad (S3)$$

where $i$ represent an arbitrary set of orthonormal single-particle states, and $|0\rangle$ is the Fock vacuum. Any SSD can be fully characterized by the set of one-body expectation values

$$\Delta_{ij} = \langle SSD | c_i^\dagger c_j | SSD \rangle \quad (S4)$$

The average of any even number of creation and destruction operators can be decomposed into a sum of products of matrix elements of $\Delta$ using Wick's theorem. For example,

$$\langle SSD | c_i^\dagger c_j^\dagger c_l c_m | SSD \rangle = \Delta_{im}\Delta_{jl} - \Delta_{il}\Delta_{jm} \quad (S5)$$

In the continuum we assume that translation symmetry is preserved upto intervalley coherence. This implies that we can characterize any HF state by,

$$\langle HF|c^\dagger_{\alpha s k} c_{\alpha' s' k'}|HF\rangle = \delta_{kk'}\Delta_{\alpha' s',\alpha s} \quad (S6)$$

As seen in Eq. S7 below, the ground state energy of an arbitrary state in the HF approximation depends only on six couplings: The energy corresponding to each $v_\mu(\mathbf{q})$ depends only on the Hartree number $g_{\mu,H} = \frac{v_\mu(\mathbf{0})}{2\pi l^2}$, and the Fock number $g_{\mu,F} = \int \frac{d\mathbf{q}}{(2\pi)^2} v_\mu(\mathbf{q}) e^{-q^2 l^2/2}$. Ignoring terms containing $g_{0,H}$, $g_{0,F}$, whose contributions to the energy are independent of all order parameters, we obtain the general expression for the HF ground state energy per guiding center

$$\frac{\mathcal{E}}{N_\phi} = -E_Z Tr(\Delta \sigma_z) + \frac{g_{z,H}}{2}(Tr(\Delta\tau_z))^2 + \frac{g_{xy,H}}{2}\left[(Tr(\Delta\tau_x))^2 + (Tr(\Delta\tau_y))^2\right] - \frac{g_{z,F}}{2}\sum_{\alpha\beta ss'}(-1)^{\alpha+\beta}\Delta^{ss'}_{\alpha\beta}\Delta^{s's}_{\beta\alpha}$$
$$- 2g_{xy,F}\sum_{ss'}\Delta^{ss'}_{KK}\Delta^{s's}_{K'K'}. \quad (S7)$$

At charge neutrality, $\nu = 0$, we must have two superpositions of the four available states at each guiding center occupied. We will assume the two occupied states to conform to the ansatz mentioned in the main text, which we repeat here again for simplicity.

$$|a\rangle = \tfrac{1}{\sqrt{2}}\left(c_a|K\uparrow\rangle - s_a|K\downarrow\rangle + c_a|K'\uparrow\rangle + s_a|K'\downarrow\rangle\right) \quad (S8a)$$
$$|b\rangle = \tfrac{1}{\sqrt{2}}\left(-c_b|K\uparrow\rangle + s_b|K\downarrow\rangle + c_b|K'\uparrow\rangle + s_b|K'\downarrow\rangle\right) \quad (S8b)$$

where $c_i = \cos\psi_i/2$ and $s_i = \sin\psi_i/2$. The form of the $\Delta$ matrix is simply the sum of the projectors of the two states $\Delta = |a\rangle\langle a| + |b\rangle\langle b|$. Explicitly, it is

$$\Delta = \frac{1}{4}\begin{pmatrix} 2+\cos\psi_a+\cos\psi_b & -\sin\psi_a-\sin\psi_b & \cos\psi_a-\cos\psi_b & \sin\psi_a-\sin\psi_b \\ -\sin\psi_a-\sin\psi_b & 2-\cos\psi_a-\cos\psi_b & \sin\psi_b-\sin\psi_a & \cos\psi_a-\cos\psi_b \\ \cos\psi_a-\cos\psi_b & \sin\psi_b-\sin\psi_a & 2+\cos\psi_a+\cos\psi_b & \sin\psi_a+\sin\psi_b \\ \sin\psi_a-\sin\psi_b & \cos\psi_a-\cos\psi_b & \sin\psi_a+\sin\psi_b & 2-\cos\psi_a-\cos\psi_b \end{pmatrix} \quad (S9)$$

This gives the following expression for the HF ground state energy.

$$\frac{\mathcal{E}}{N_\phi} = \frac{1}{2}\left(\cos^2\psi_a + \cos^2\psi_b\right)(g_{xy,H} - g_{xy,F})$$
$$- \frac{1}{2}\cos(\psi_a+\psi_b)(g_{xy,F} + g_{xy,H})$$
$$- \frac{1}{2}\cos(\psi_a-\psi_b)(g_{z,F} + g_{xy,H})$$
$$- \frac{1}{2}(g_{xy,F} + g_{z,F}) - E_z(\cos\psi_a + \cos\psi_b) \quad (S10)$$

The minimization conditions are,

$$\frac{\partial\mathcal{E}}{\partial\psi_a} = 0 = \frac{1}{2}\sin(\psi_a+\psi_b)(g_{xy,F}+g_{xy,H})$$
$$+ \frac{1}{2}\sin(\psi_a-\psi_b)(g_{z,F}+g_{xy,H})$$
$$- \sin\psi_a\cos\psi_a(g_{xy,H}-g_{xy,F})$$
$$+ E_z\sin\psi_a \quad (S11)$$

$$\frac{\partial\mathcal{E}}{\partial\psi_b} = 0 = \frac{1}{2}\sin(\psi_a+\psi_b)(g_{xy,F}+g_{xy,H})$$
$$- \frac{1}{2}\sin(\psi_a-\psi_b)(g_{z,F}+g_{xy,H})$$
$$- \sin\psi_b\cos\psi_b(g_{xy,H}-g_{xy,F})$$
$$+ E_z\sin\psi_b \quad (S12)$$

One needs to solve this set of equations for all values of couplings. While we have done this numerically, there is an elegant way to find the phase boundaries, which reproduces those found numerically. In the region of interest, where physical samples are believed to be, we have two known solutions. One is the CAF, with $\psi_a = \psi_b = \theta$, with the optimal value of $\theta$ being determined by $E_Z$ for $g_{xy,F} < 0$ and $E_Z < 2|g_{xy,F}|$ as

$$\cos\theta = \frac{E_Z}{2|g_{xy,F}|} \quad (S13)$$

The HF ground state energy of the CAF solution is

$$\frac{\mathcal{E}_{CAF}}{N_\phi} = -g_{z,F} - \frac{E_Z^2}{2|g_{xy,F}|} \quad (S14)$$

The other solution of interest is the BO, with $\psi_a =$

$0, \psi_b = \pi$, and HF ground state energy

$$\frac{\mathcal{E}_{BO}}{N_\phi} = 2g_{xy,H} - g_{xy,F} \tag{S15}$$

In the model with ultra-short-range interactions, these two states are separated by a first-order transition. However, we wish to know whether one can have a coexistence of CAF and BO. Generically, we expect that coexistent phases will be separated by second-order transitions. At a second-order transition from the CAF to the phase in which BO coexists with CAF order, we expect an instability of the CAF towards BO within the two-angle ansatz. Since each solution is a local extremum, we need to examine the matrix of second derivatives of the HF ground state energy in the space of the two angles $\psi_a, \psi_b$. If this matrix has both eigenvalues positive, the solution is stable. A zero crossing of an eigenvalue signals an instability of the state. First let us examine the matrix of second derivatives at the CAF state.

$$\begin{bmatrix} \frac{\partial^2 \mathcal{E}}{\partial \psi_a^2} & \frac{\partial^2 \mathcal{E}}{\partial \psi_a \partial \psi_b} \\ \frac{\partial^2 \mathcal{E}}{\partial \psi_b \partial \psi_a} & \frac{\partial^2 \mathcal{E}}{\partial \psi_b^2} \end{bmatrix}_{\text{CAF}} \tag{S16}$$

Explicitly, we find,

$$\left.\frac{\partial^2 \mathcal{E}}{\partial \psi_a^2}\right|_{\text{CAF}} = \frac{3}{2}\cos 2\theta g_{xy,F} + \frac{g_{z,F}}{2} + E_z \cos\theta$$
$$- \frac{1}{2}\cos 2\theta g_{xy,H} + \frac{g_{xy,H}}{2} \tag{S17}$$

$$\left.\frac{\partial^2 \mathcal{E}}{\partial \psi_b^2}\right|_{\text{CAF}} = \frac{3}{2}\cos 2\theta g_{xy,F} + \frac{g_{z,F}}{2} - \frac{1}{2}\cos 2\theta g_{xy,H}$$
$$+ \frac{g_{xy,H}}{2} + E_z \cos\theta \tag{S18}$$

$$\left.\frac{\partial^2 \mathcal{E}}{\partial \psi_a \partial \psi_b}\right|_{\text{CAF}} = \frac{1}{2}\cos 2\theta \left(g_{xy,F} + g_{xy,H}\right)$$
$$- \frac{1}{2}\left(g_{z,F} + g_{xy,H}\right) \tag{S19}$$

One can see from here that we have $\left.\frac{\partial^2 \mathcal{E}}{\partial \psi_a^2}\right|_{\text{CAF}} = \left.\frac{\partial^2 \mathcal{E}}{\partial \psi_b^2}\right|_{\text{CAF}}$.

Thus, the eigenvalues are $\left.\frac{\partial^2 \mathcal{E}}{\partial \psi_a^2}\right|_{\text{CAF}} \pm \left.\frac{\partial^2 \mathcal{E}}{\partial \psi_a \partial \psi_b}\right|_{\text{CAF}}$.

For the simple, though physically unrealistic, case $E_z = 0$, we obtain the condition for the instability as,

$$g_{z,F} = 2|g_{xy,H}| - |g_{xy,F}|, \text{ where } g_{xy,H}, g_{xy,F} < 0 \tag{S20}$$

This is precisely the condition for the HF ground state energies to cross, and so does not allow for coexistence.

Thus, one does not have coexistence between the CAF and BO phases when the Zeeman coupling vanishes, even with non-zero-range interactions.

Now we go to the physically realistic case $E_z > 0$, obtaining the condition for an instability of the CAF state to be,

$$g_{z,F} - 2|g_{xy,H}| + |g_{xy,F}| + \frac{E_z^2}{2|g_{xy,F}|}$$
$$+ \frac{E_z^2}{2|g_{xy,F}|^2}\left(|g_{xy,H}| - |g_{xy,F}|\right) = 0 \tag{S21}$$

Let us compare this to the difference of the two ground state energies.

$$\frac{\mathcal{E}_{BO} - \mathcal{E}_{CAF}}{N_\phi} = g_{z,F} - 2|g_{xy,H}| + |g_{xy,F}| + \frac{E_z^2}{2|g_{xy,F}|} = 0 \tag{S22}$$

Thus when $\left(|g_{xy,H}| - |g_{xy,F}|\right) < 0$, the BO state is higher than the CAF state in energy, and yet there is an instability of the CAF state. This indicates an intermediate state which manifests the coexistence of the two order parameters (see fig. S1).

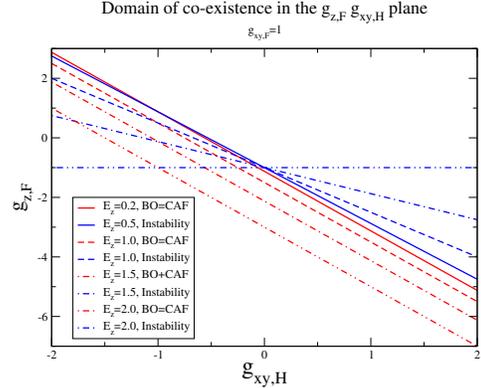

FIG. S1. Domain Coexists

Thus, the two necessary conditions for the coexistence of CAF and BO are $0 > g_{xy,H} > g_{xy,F}$ and $E_Z > 0$.

## SIII. RENORMALIZATION OF INTERACTIONS

In the previous section we saw that in order to have coexistence between the CAF and BO order parameters, one needs $0 > g_{xy,H} > g_{xy,F}$ and $2|g_{xy,F}| > E_Z > 0$. A significant difference between $g_{xy,H}$ and $g_{xy,F}$ can only arise if $v_{xy}(\mathbf{q})$ has nontrivial structure on the scale of $\ell$. In this section we present an illustrative calculation demonstrating that integrating out the higher Landau levels can indeed introduce structure at the length scale $\ell$ into the effective interactions.

Let us start with the interacting Hamiltonian in the space of all Landau Levels in the MLG. This is not actually the "correct" interaction [1], because it ignores form factors arising from the sublattice degree of freedom, but is an illustrative example that satisfies all the symmetries.

$$H = \sum_{n\alpha sk} \omega_c \sqrt{|n|} c^\dagger_{n\alpha sk} c_{n\alpha sk} + \frac{1}{2L_x L_y} \sum_{\substack{\mathbf{q} k_1 k_2 \\ n_i m_i}} \rho_{n_1 n_2}(\mathbf{q}) \rho_{m_1 m_2}(-\mathbf{q}) e^{-iq_x(k_1-k_2-q_y)l^2} \times$$

$$\left[ v_0(\mathbf{q}) \sum_{\substack{\alpha\beta \\ s_1 s_2}} c^\dagger_{n_1\alpha s_1 k_1-q_y} c^\dagger_{m_1\beta s_2 k_2+q_y} c_{m_2\beta s_2 k_2} c_{n_2\alpha s_1 k_1} + v_z(\mathbf{q}) \sum_{\substack{\alpha\beta \\ s_1 s_2}} (-1)^{\alpha+\beta} c^\dagger_{n_1\alpha s_1 k_1-q_y} c^\dagger_{m_1\beta s_2 k_2+q_y} c_{m_2\beta s_2 k_2} c_{n_2\alpha s_1 k_1} \right.$$

$$\left. + J(\mathbf{q}) \sum_{\substack{\alpha\beta \\ s_i s'_i \\ a}} (\sigma^a)_{s_1 s'_1} (\sigma^a)_{s_2 s'_2} c^\dagger_{n_1\alpha s_1 k_1-q_y} c^\dagger_{m_1\beta s_2 k_2+q_y} c_{m_2\beta s'_2 k_2} c_{n_2\alpha s'_1 k_1} \right] \quad (S23)$$

Note that since we wish to integrate out higher LLs, we need to introduce the LL indices $m, n$. The first term encodes the Dirac-Landau spectrum of graphene.

We rewrite the interaction terms for future convenience as,

$$\frac{1}{2L_x L_y} \sum_{s_i \alpha_i} V^{\{s_i\}}_{\{\alpha_i\}}(\mathbf{q}) \rho_{m_1 m_4}(\mathbf{q}) \rho_{m_2 m_3}(-\mathbf{q}) e^{-iq_x(k_1-k_2-q_y)l^2}$$

$$\times c^\dagger_{m_1\alpha_1 s_1 k_1-q_y} c^\dagger_{m_2\alpha_2 s_2 k_2+q_y} c_{m_3\alpha_3 s_3 k_2} c_{m_4\alpha_4 s_4 k_1} \quad (S24)$$

Integrating out states is easiest in a path integral formulation. We therefore express the problem in the imaginary time path integral replacing the operators $c_{m\alpha sk}$ by the Grassmann numbers $\psi_{m\alpha s}(k,\tau)$. The interacting part of the action is

$$S_{\text{int}} = \frac{1}{2L_x L_y} \int \frac{d\omega_1 d\omega_2 d\omega_3}{(2\pi)^3} \sum_{\substack{m_i s_i \\ \alpha_i k_i \\ \mathbf{q}}} V^{s_i}_{\alpha_i}(\mathbf{q})$$

$$\rho_{m_1 m_4}(\mathbf{q}) \rho_{m_2 m_3}(\mathbf{q}) e^{-iq_x(k_1-k_2-q_y)l^2}$$
$$\bar\psi_{m_1\alpha_1 s_1}(k_1-q_y,\omega_1) \bar\psi_{m_2\alpha_2 s_2}(k_2+q_y,\omega_2)$$
$$\psi_{m_3\alpha_3 s_3}(k_2,\omega_3) \psi_{m_4\alpha_4 s_4}(k,\omega_1+\omega_2-\omega_3) \quad (S25)$$

The free propagator is,

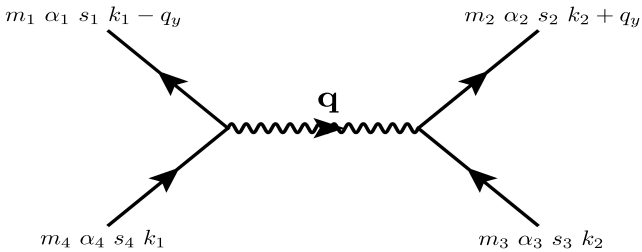

FIG. S2. Fundamental Vertex

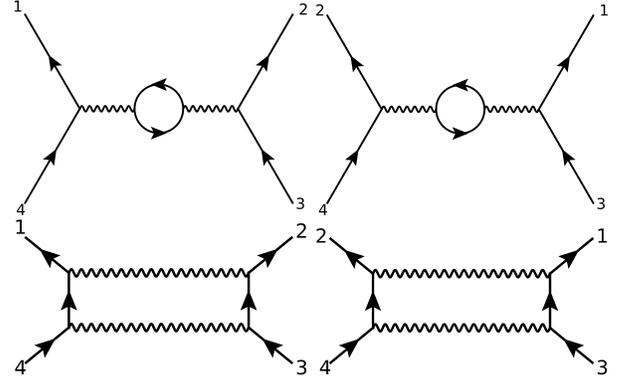

FIG. S3. Main Diagrams

$$\langle \bar\psi_{m_1\alpha_1 s_1}(k_1,\omega_1) \psi_{m_2\alpha_2 s_2}(k_2,\omega_2) \rangle =$$
$$\frac{\delta_{m_1 m_2} \delta_{\alpha_1\alpha_2} \delta_{s_1 s_2} \delta_{k_1 k_2} 2\pi \delta(\omega_1-\omega_2)}{i\omega_1 - \mathcal{E}_{m_1,\alpha_1,s_1}} \quad (S26)$$

where the 1-body energy $\mathcal{E}_{m_1,\alpha_1,s_1}$ will, in general, also get self-energy corrections as the states are integrated out.

Our goal in this section is to provide a proof of principle that effective interactions in the $n=0$ manifold of LLs in graphene can acquire structure at the length scale $\ell$, rather than performing an accurate computation. Thus, we will ignore all self-energy corrections, replacing the 1-body energies by the non-interacting Dirac-Landau spectrum.

As usual in Wilsonian renormalization group, we separate states into a slow, or low-energy sector ($<$) and a fast, or high-energy sector ($>$). We integrate out the high-energy sector, schematically obtaining,

$$\left\langle e^{-S_{\text{int}}} \right\rangle_> = e^{-\langle S_{\text{int}} \rangle_> + \frac{1}{2}\langle S^2_{\text{int}} \rangle_{c,>} + \ldots} \quad (S27)$$

$$\Rightarrow \delta S = -\frac{1}{2} \left\langle S^2_{\text{int}} \right\rangle_> \quad (S28)$$

where the average is taken over fast modes. There are four different diagrams S3 that will contribute to the fun-

damental vertex S2. The easiest to analyze is S4 because this is the only diagram resummed in RPA and also the only diagram that gives back a density-density form. The

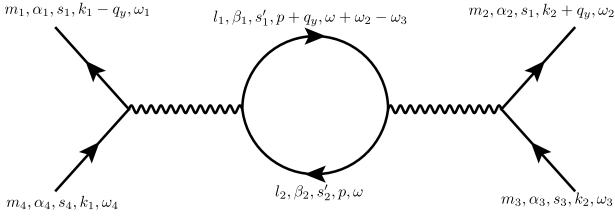

FIG. S4. Convention of momentum and other indices where $\omega_4 = \omega_1 + \omega_2 - \omega_3$.

contribution of this diagram is,

$$\sum_{\substack{l_1,l_2 \\ \beta_1,\beta_2 \\ s'_1,s'_2,p}} \int_\infty^\infty \frac{d\omega}{4\pi \left[i\omega - \mathcal{E}_{l_2,\beta_2,s'_2}\right]\left[i(\omega+\omega_2-\omega_3) - \mathcal{E}_{l_1,\beta_1,s'_1}\right]}$$

$$\times \frac{1}{4L_x^2 L_y^2} \sum_{q_x q'_x q_y} V^{s_1 s'_1 s'_2 s_4}_{\alpha_1 \beta_1 \beta_2 \alpha_4}(q_x, q_y) V^{s'_2 s_2 s_3 s'_1}_{\beta_2 \alpha_2 \alpha_3 \beta_1}(q'_x, q_y)$$

$$\times e^{-iq_x(k_1-p-q_y)l^2} e^{-iq'_x(p+q_y-k_2-q_y)l^2} \rho_{m_1 m_4}(q_x, q_y)$$

$$\times \rho_{l_1 l_2}(-q_x,-q_y) \rho_{l_2 l_1}(q'_x, q_y) \rho_{m_2 m_3}(-q'_x, -q_y).$$
(S29)

Since $\mathcal{E}$ is independent of $p$ we can perform the sum over $p$, $\sum_p e^{i(q_x - q'_x)p} = \frac{L_x L_y}{2\pi l^2} \delta_{q_x, q'_x}$ Thus,

$$\delta S_{\text{int}} = \sum_{\substack{\alpha_i, \beta_i \\ m_i, l_i \\ s_i, \mathbf{q}}} \frac{1}{8\pi l^2 L_x L_y} e^{-iq_x(k_1-k_2-q_y)l^2}$$

$$\times V^{s_1 s'_1 s'_2 s_4}_{\alpha_1 \beta_1 \beta_2 \alpha_4}(\mathbf{q}) V^{s'_2 s_2 s_3 s'_1}_{\beta_2 \alpha_2 \alpha_3 \beta_1}(\mathbf{q})$$

$$\times \rho_{l_1,l_2}(\mathbf{q}) \rho_{l_2,l_1}(-\mathbf{q}) \rho_{m_1,m_2}(\mathbf{q}) \rho_{m_3,m_4}(-\mathbf{q})$$

$$\times \frac{n_F(l_1, s'_1) - n_F(l_2, s'_2)}{\mathcal{E}_{l_1, s'_1} - \mathcal{E}_{l_2, s'_2}} \quad (S30)$$

Now we have $\rho^*_{l_1, l_2}(\mathbf{q}) = \rho_{l_2,l_1}(-\mathbf{q})$ and for non-interacting graphene,

$$\mathcal{E}_{l,s} = \text{sgn}(l)\omega_c\sqrt{|l|} \quad (S31)$$

In order for the $\omega$ integral to be nonvanishing, one of the $l_1, l_2$ must be positive and the other negative, say $l_1 > 0$ and $l_2 < 0$. So, the renormalization from integrating out $l_1, l_2$ is,

$$\delta V^{s_1 s_2 s_3 s_4}_{\alpha_1 \alpha_2 \alpha_3 \alpha_4}(\mathbf{q}) = -\frac{1}{2} \sum_{\substack{s'_1 s'_2 \\ \beta_1 \beta_2}} \left|\rho_{l_1 l_2}(\mathbf{q})\right|^2 V^{s_1 s'_1 s'_2 s_4}_{\alpha_1 \beta_1 \beta_2 \alpha_4}(\mathbf{q})$$

$$\times V^{s'_2 s_2 s_3 s'_1}_{\beta_2 \alpha_2 \alpha_3 \beta_1}(\mathbf{q}) \frac{1}{\omega_c\left(\sqrt{|l_1|} + \sqrt{|l_2|}\right)}$$
(S32)

This clearly depends on the magnetic length via $\left|\rho_{l_1 l_2}(\mathbf{q})\right|^2$.

There are other diagrams which we have not computed, which will also renormalize the effective interactions, which in general will not have the density-density form. These will also bring in the length scale $\ell$. It is highly implausible that the structure at the length scale $\ell$ would cancel between the different contributions. Thus, we conclude that, generically, contributions to the effective interactions in the $n = 0$ manifold of graphene will have structure on the scale of the magnetic length.

## SIV.  LATTICE MODEL

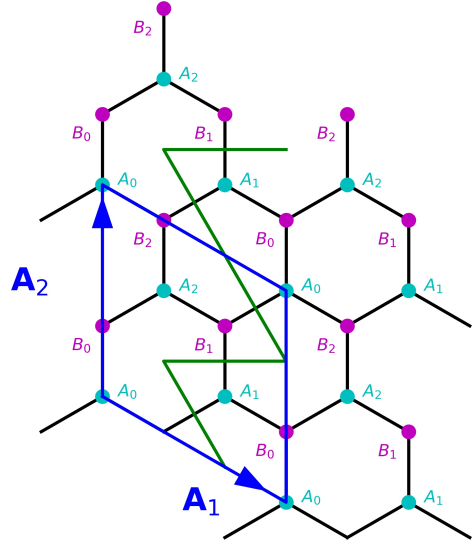

FIG. S5. We present here the Kekulé unit cell (KUC) and the gauge choice for our calculations. The blue arrows in the picture shows the two vectors $\mathbf{A}_1, \mathbf{A}_2$ that defines the KUC. The KUC is made out of 3 unit cells hence has 6 sublattice sites as indicated by $A_\alpha, B_\alpha$ where $\alpha \in \{1, 2, 3\}$. We also present here a way to choose the phases on the bonds for this unit cell. The green line is drawn in a way that it can go from one KUC to next, travelling through all the honeycombs exactly once. Every time it crosses a bond we choose that phase of the hopping on that bond to be the phase on the previous crossed bond plus $\phi_0/q$ ($\phi_0 = h/w$ being flux quantum) keeping the direction of the positive phase the same with respect to the green line. Thus when $q$ a multiple of 3, the effective unit cell with the smallest integer number of flux quanta will just be $q$ times larger than the conventional unit cell for graphene. For $q$ not a multiple of 3 that will be $3q$ times larger than the conventional unit cell.

We know in the presence of $1/q$ flux quantum per original unit cell, we need a unit cell at least $q$ times larger. This is because we want two commuting translations, which can only occur in a larger unit cell which is pierced by an integer number of flux quanta. This larger unit cell

is called the magnetic unit cell (MUC). We define an extended unit cell (see Fig. S5) such that it *can fit the Kekulé bond order* state. The primitive lattice vectors that define the original unit cell of graphene are

$$\mathbf{a}_1 = \hat{x} \tag{S33a}$$
$$\mathbf{a}_2 = \frac{1}{2}\hat{x} + \frac{\sqrt{3}}{2}\hat{y}. \tag{S33b}$$

The unit cell that can fit the Kekulé distortion will be defined with the lattice vectors $\mathbf{A}_1 = 2\mathbf{a}_1 - \mathbf{a}_2$ and $\mathbf{A}_2 = 2\mathbf{a}_2 - \mathbf{a}_1$. One can easily see that the area enclosed by $\mathbf{A}_1, \mathbf{A}_2$ is three times larger than the area of the primitive unit cell of graphene. For brevity we will call this unit cell the KUC. We also choose a gauge for this unit cell as described in the caption of Fig. S5. For $q$ a multiple of 3, the MUC will be a simple multiple of KUC in the $\mathbf{A}_2$ direction and the MUC will be $q$ times larger than the primitive cell made out of $q/3$ KUCs. For $q$ not a multiple of 3, $q \times$KUC defines the MUC for our calculation.

Our model is represented by the following Hamiltonian

$$\begin{aligned}
H = &- E_z \sum_{\mathbf{R}_i} S_x(\mathbf{R}_i) \\
&- \sum_{\langle \mu,\mathbf{R}_i;\nu,\mathbf{R}_j\rangle,s} e^{i\phi_{\mu,\nu}} c^\dagger_{\mu,s}(\mathbf{R}_i) c_{\nu,s}(\mathbf{R}_j) \\
&+ \frac{U}{2} \sum_{\mathbf{R}_i,\mu,s\neq s'} :n_{\mu,s}(\mathbf{R}_i) n_{\mu,s'}(\mathbf{R}_i): \\
&- 2g \sum_{\langle \mu,\mathbf{R}_i;\nu,\mathbf{R}_j\rangle} :\mathbf{S}_\mu(\mathbf{R}_i)\cdot\mathbf{S}_\nu(\mathbf{R}_j):
\end{aligned} \tag{S34}$$

where $\mu,\nu$ are sublattice indices, $\mathbf{R}_i, \mathbf{R}_j$ are MUC sites, and $s, s'$ are spin indices. We define the operators $S^\alpha_\mu(\mathbf{R}_i) = \frac{1}{2}\sum_{s,s'} c^\dagger_{\mu,s}(\mathbf{R}_i) \sigma^\alpha_{s,s'} c_{\mu,s'}(\mathbf{R}_i)$ and $n_{\mu,s'}(\mathbf{R}_i) = c^\dagger_{\mu,s}(\mathbf{R}_i) c_{\mu,s}(\mathbf{R}_i)$. The phase of the hopping $\phi_{\mu,\nu}$ is chosen as described in Fig. S5. The notation $:\mathcal{O}:$ is for normal ordering of the operator string $\mathcal{O}$ with respect to Fock vacuum (defined as all creation operators on the left and all annihilation operators on the right).

To analyze our model Hamiltonian (Eq. S34) we will use self-consistent Hartree-Fock theory. We will only study the phases for half filling which corresponds to $\nu = 0$. To perform the self-consistent HF method we will use the following assumptions.

1. We will assume the magnetic crystal momenta defined with respect to the MUC are unbroken in HF. Thus for HF ground state $|\Omega\rangle$ we can write,

$$\langle\Omega|C^\dagger_\beta(\mathbf{k})C_\alpha(\mathbf{k}')|\Omega\rangle = \delta_{k,k'}\Delta_{\alpha,\beta}(\mathbf{k}), \tag{S35}$$

Then we minimize the energy with respect to $\Delta_{\alpha,\beta}(\mathbf{k})$.

2. We will allow all possible mixings of sublattice and spin inside the MUC.
3. We will assume the bands are smooth and we can sample the BZ at a few representative momentum points to evaluate $\Delta$ and the energy.

Here we present the phase diagram (Fig. S6 for $q = 6$ at $E_z = 0.25$. As can be seen, BO coexists not only with CAF but also with charge density wave (CDW). There are several bond-ordered phases with no other order parameter, and with different kinds of bond order. The BO that coexists with the CAF is similar to Kekulé order, but the others are quite different. Except for the ferromagnetic phase we find BO to be ubiquitous, occurring either alone or in coexistence with other order parameters.

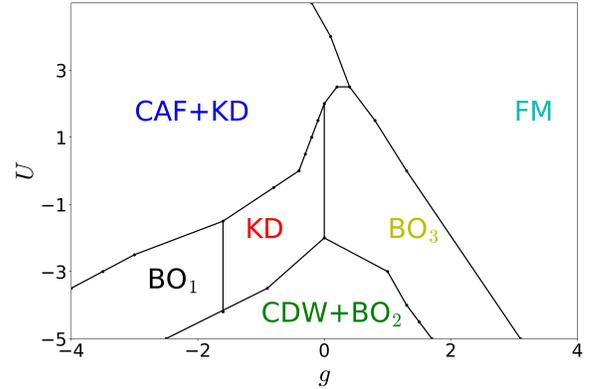

FIG. S6. Lattice Phase Diagram for $q = 6$ at $E_z = 0.25$. We see phases with different kinds of bond-order in different parts of the phase diagram. The CAF and CDW order parameters always coexist with bond-order. The Ferromagnet is the sole state that shows no coexistence.

[1] B. Feshami and H. A. Fertig, Hartree-fock study of the $\nu = 0$ quantum hall state of monolayer graphene with short-range interactions, Phys. Rev. B **94**, 245435 (2016).



\end{document}